\documentclass[twocolumn,trackchanges]{aastex631}

\hypersetup{linkcolor=red,citecolor=green,filecolor=cyan,urlcolor=magenta}
\usepackage{comment}
\usepackage{hyperref}
\usepackage{amsmath}
\usepackage{longtable}
\usepackage{longfigure}

\usepackage{supertabular}

\begin{document}

\title{Constraining X-ray variability of the blazar 3C 273 using \textit{XMM-Newton} observations over two decades}

\author{Adithiya Dinesh}
\affiliation{IPARCOS Institute and EMFTEL Department, Universidad Complutense de Madrid, E-28040 Madrid, Spain}

\author{Gopal Bhatta}
\affiliation{Janusz Gil Institute of Astronomy, University of Zielona G\'ora, ul. Szafrana 2, 65-516 Zielona G\'ora, Poland}


\author{Tek P. Adhikari}
\affiliation{Pokhara Astronomical Society (PAS), Bhajapatan -13, Pokhara, Nepal}
\affiliation{Nicolaus Copernicus Astronomical Center, Bartycka 18, 00-716, Warsaw, Poland}

\author{Maksym Mohorian}
\affiliation{School of Mathematical and Physical Sciences, Macquarie University, Sydney, NSW 2109, Australia}
\affiliation{Astronomy, Astrophysics and Astrophotonics Research Centre, Macquarie University, Sydney, NSW 2109, Australia}

\author{Niraj Dhital}
\affiliation{Central Department of Physics, Tribhuvan University, Kirtipur 44613, Nepal}

\author{Suvas C. Chaudhary}
\affiliation{Inter-University Centre for Astronomy and Astrophysics, Pune, Maharashtra 411007, India}

\author{Radim P{\'a}nis}
\affiliation{Research Centre for Theoretical Physics and Astrophysics, Institute of Physics,\\
Silesian University in Opava, Bezru{\v c}ovo n{\'a}m.13, CZ-74601 Opava, Czech Republic}

\author{Dariusz G\'ora}
 \affiliation{Institute of Nuclear Physics, Polish Academy of Sciences, 
 PL-31342 Krak\'ow, Poland}

\begin{abstract}
Blazars exhibit relentless variability across diverse spatial and temporal frequencies. The study of long- and short-term variability properties observed in the X-ray band provides insights into the inner workings of the central engine. In this work, we present timing and spectral analyses of the blazar 3C 273 using the X-ray observations from the $\textit{XMM-Newton}$ telescope covering the period from 2000 to 2020. The methods of timing analyses include estimation of fractional variability, long- and short-term flux distribution, rms-flux relation, and power spectral density analysis.  The spectral analysis include estimating a model independent flux hardness ratio and fitting the observations with multiplicative and additive spectral models such as \textit{power-law}, \textit{log-parabola}, \textit{broken power-law}, and \textit{black body}. The \textit{black body} represents the thermal emission from the accretion disk, while the other models represent the possible energy distributions of the particles emitting synchrotron radiation in the jet. During the past two decades, the source flux changed by of a factor of three, with a considerable fractional variability of 27\%. However, the intraday variation was found to be moderate. Flux distributions of the individual observations were consistent with a normal or log-normal distribution, while the overall flux distribution including entire observations appear to be rather multi-modal and of a complex shape. The spectral analyses indicate that \textit{log-parabola} added with a \textit{black body} gives the best fit for most of the observations. The results indicate a complex scenario in which the variability can be attributed to the intricate interaction between the disk/corona system and the jet.
\end{abstract}


\section{Introduction} \label{sec:intro}
Blazars are Active Galactic Nuclei (AGN) that have a relativistic jet directed towards our line of sight \citep{urry1995unified}. These are highly energetic objects in the Universe with characteristics such as rapid flux variations, high polarization, and apparent superluminal motion \citep{jorstad2001multiepoch,marscher2008core}.
 The observed X-ray emission from these sources is mostly from their
jets due to relativistic beaming, which amplifies the apparent emission from the jet in the observer's frame.
 Blazars emit across the complete electromagnetic spectrum, which
is predominantly non-thermal in nature. Highly variable linear polarisation has also been detected in several sources in the radio to optical regime \citep{villforth2010variability, Bhatta2015,falomo2014optical}.
 
The spectral energy distribution (SED) of blazars has a distinctive two-humped shape, in the $\nu$-$F\nu$ diagram, where $F\nu$ is the flux density (ergs\ cm$^{-2}$\ s$^{-1}$\ Hz$^{-1}$) and $\nu$ is the frequency. 
The lower-frequency hump is generally associated with synchrotron radiation from the relativistic particles moving in the magnetic field of the jets; however, the origin of the higher-frequency hump is still debated. It is commonly agreed that this feature results via inverse-Compton scattering of low frequency photons by relativistic particles. There are mainly two proposed mechanisms to explain the origin of such emission in the high frequencies. In the leptonic models of blazars, the primary particles responsible for the high energy emission are electrons or positrons and the high energy emission results when the relativistic electrons in the jets inverse-Compton scatter the low frequency photons. In the synchrotron self-Compton (SSC) model, the soft energy  seed photons are mainly contributed by the synchrotron emission \citep{maraschi1992jet}; whereas in the external Compton model
\citep{dermer1993model, sikora1994comptonization} the seed photons can be contributed by accretion disk, broad line region (BLR) and dusty torus. On the other hand, in hadronic models of blazars the observed high energy emission primarily results due to interaction of protons in the jets \citep{aharonian2000tev,mucke2003bl}; and synchrotron radiation is produced due to the secondary particles from the proton-proton interactions \citep{mucke2001proton,cerruti2015hadronic}.

Blazars are broadly classified in two main subgroups: flat spectrum radio quasars (FSRQ), the more powerful type that shows optical emission lines above the continuum and BL Lacerate objects (BL Lac), the less powerful sources showing weak or absence of emission lines in their featureless continuum \textit{power-law} emission. In addition, a subdivision scheme based on the location of the synchrotron peak in the $\nu$-$F\nu$ plane is used to define low synchrotron peaked blazars ($\nu_{\rm s} \leq 10^{14.0}$ Hz), intermediate synchrotron peaked blazars ($10^{14}<\nu_{\rm s} \leq10^{15} $ Hz), and high synchrotron peaked blazars ($\nu_{\rm s} > 10^{15}$ Hz) (\citealt{Fan2016}, see also \citealt{Abdo2010}).
Therefore, FSRQs are Compton dominant powerful sources with the synchrotron peak in the lower frequency, whereas BL Lac objects, although relatively less powerful, are the TeV sources with the inverse-Compton peak in the highest $\gamma$-ray bands. 
As mentioned, blazars are highly variable sources, with the variability timescales spanning from a few minutes to several decades. Variability over time scales from a few minutes to less than a day is often termed intraday or micro-variability, from days to weeks is termed as short timescale variability, and from months to years is long-term variability \citep[see, e.g.,][]{wagner1995intraday,Bhatta2021,Webb2021}. Blazars represent some of the most luminous types of astrophysical sources that host extreme physical conditions around super-massive black holes, leading to the ejection of relativistic jets and the production of highly energetic $\gamma$-ray emission. The central engine of the sources is still unresolved by the current instrument; in such contexts, variability analysis can be used as a powerful tool to probe the nature of the central engine and emission mechanism of blazars. Studying variability helps in constraining the size, magnetic field and other properties around the compact region in blazars. In particular, as X-rays can penetrate deeper into the innermost regions, a study of X-ray flux and spectral variability provides an excellent probe into the violent episodes in both the accretion disk and the jets. In general, the observed variability is aperiodic in nature, such that its statistical properties can be well represented by \textit{power-law} type Power Spectral Density (PSD). But in some sources, the light curves exhibit quasi-periodic oscillations in the flux \citep[see, e.g.,][]{sillanpaa1988oj,Bhatta2019}.

 In this work, we analyze the archival data from $\textit{XMM-Newton}$ spanning the period from 2000 to 2020 to conduct an extensive variability study of source 3C 273. By utilizing the light curve with a long-time baseline and employing analyses that combine both timing and spectral methods, we aim to constrain the physical mechanisms at the central engine of the AGN that shape the observed intra-day and long-term variability in the X-ray band.

The paper is organized as follows. In Section \ref{sec:2}, we present a brief introduction about source 3C 273. We discuss the data reduction methods in Section \ref{sec:3}. Spectral and variability analysis is presented in Section \ref{sec:4}. We present the results and discussion in Section \ref{sec:5} and the conclusions in Section \ref{sec:6}.

\section{Blazar 3C 273} \label{sec:2}
Blazar 3C 273 is the first discovered quasar \citep[see, e.g.,][]{edge1959survey,schmidt19633c}. It is classified as an FSRQ and is luminous in all frequencies \citep{schmidt19633c}. It is also the brightest quasar and is located at a distance $z$ = 0.158. Its emission is dominated by the synchrotron emission from the jet \citep{turler2000modelling} and it has a radio core which exhibits superluminal disturbances \citep{jorstad2012parsec}.

3C 273 is an extensively studied source across the electromagnetic spectrum. \citet{courvoisier1998bright} provides a review of the literature on 3C 273 till 1998. \citet{sambruna2001chandra} studied the jet in X-rays, where inverse Compton scattering was found to explain the SED at some regions. \citet{haardt2008hidden} studied the properties of 3C 273 like the fluorescent K$\alpha$ transition line, the absorption edge, and the soft excess and came to the conclusion that these are variable in different observations over different periods and might be linked to the source's state. The variable soft content in the X-ray emission of this source was observed and studied extensively (\cite{courvoisier1987radio}, \cite{turner1990x}, \cite{leach1995constraining}, \cite{turler2006historic}, \cite{soldi2008multiwavelength}, \cite{pietrini2008possible}). Another interesting feature present in 3C 273 is a large optical/ultraviolet bump \citep{paltani1998blue}. It is thought to be the result of accretion disc emission. In a more recent cross-correlation study using decade-long light curves of blazars, the source was found to show little correlation between the optical and $\gamma$-ray, which can be interpreted as the optical emission mostly arising from the accretion disk \citep[see, e.g.,][]{Bhatta2021}.

The X-ray and IR emission of 3C 273 and their correlation were studied by \cite{mchardy2007simultaneous}. Also, studies of optical to X-ray and $\gamma$-ray emission of this source have been carried out \citep{courvoisier2003simultaneous, kalita2015multiband}. \citet{page2004xmm} studied the $\textit{XMM-Newton}$ observations from 2000 to 2003 and found that the spectra below 2 keV could be explained by either multiple black bodies or a Comptonization model. The spectra around 3–10 keV were well-fitted with a \textit{power-law} (\textit{PL}) model, sometimes along with a Fe emission line. \cite{foschini2006xmm} studied the spectral features of the source using $\textit{XMM-Newton}$ data for the period 2000-2004, which showed that there was an excess thermal component relative to the synchrotron component and a \textit{broken power-law} model could be used to explain the 0.4–10 keV spectra. \citet{courvoisier2003simultaneous} studied the spectra of this source above 3 keV and used a simple \textit{power-law} model to explain the spectra. \citet{kalita2017origin} analysed the source spectrum from 2000-2015, where they studied the spectral evolution of the source and the relation between the X-ray and UV emissions. \citet{bhattacharyya2020blazar} have conducted studies on the non-stationarity and flux-RMS relation of 3C 273. $\gamma$-ray variability using Fermi-LAT observations of 3C 273 has been conducted by \citep{bhatta2020nature}. Moreover, \citet{pavana2022} analysed 23 pointed \textit{XMM-Newton} observations of 3C 273 taken during 2000-2001 and studied the X-ray intraday variability and power spectral density (PSD). The authors concluded that both the particle acceleration and synchrotron cooling processes significantly contribute to the emission from this blazar. In the current work, we adopt a more comprehensive approach by combining results from various timing and spectral analyses on a long-time baseline in order to obtain a coherent picture.


\section{\textit{XMM-Newton} archival data of 3C 273}
\label{sec:3}
$\textit{XMM-Newton}$ is a European X-ray satellite mission with a huge collecting area (3$\times\sim$1500cm$^2$ at 1.5 keV) and the ability to observe a source in multiple bands.
In our work, we use the data from EPIC (European Photon Imaging Camera) PN observations of 3C 273 taken in imaging mode for getting high-quality data \citep{struder2001european}. All the observation IDs along with some basic findings from the timing analysis are listed in Table \ref{table1}.
Observation data files were downloaded from the online \textit{XMM-Newton} science archive\footnote{https://heasarc.gsfc.nasa.gov/W3Browse/all/xmmmaster.html}. The \textit{XMM-Newton} science analysis software (SAS)\footnote{"User's Guide to the XMM-Newton Science Analysis System", Issue 16.0, 2021 (ESA: XMM-Newton SOC)} version 19.0.0 was used for data processing.
There were a total of 45 observations during the time period we choose. However, some observations were not usable due to due to factors such as the absence of EPIC PN data, observation periods shorter than 10 ks, and poor image quality.
After considering all these factors, we selected a total of 26 observations for our analysis.

\subsection{Data Processing}
We followed the SAS data analysis threads for the data reduction \footnote{https://www.cosmos.esa.int/web/xmm-newton/sas-threads}. The task \textit{epproc} was used to produce the calibrated event files. In the very first step, we generated light curves in the energy range of $10-12$ keV to check for the soft proton flares. A good time interval (GTI) file, which contains the information of the good times that are free from soft proton flares, was then generated using the \textit{tabgtigen} tool. In the next step, we utilized the event list file and GTI file as input to obtain cleaned event list files using an expression ``FLAG==0 \&\& PATTERN$\leq$4” for all PN data (e. g., \cite{bhattacharyya2020blazar, kalita2015multiband}).  The choice of ``PATTERN$\leq$4" includes the single and double pattern source events maintaining a good signal-to-noise ratio. The value for count rates was above 4ct s$^{-1}$ for PN data to provide the longest exposure whilst minimizing contamination. Finally, these cleaned event lists are used to obtain source images, which can be further utilized to extract light curves and spectra needed for our science goals.

The background regions are extracted from nearby regions, located within a distance of   $\sim$400'' from the source regions and on the same chip. Using these regions and specific selection criteria outlined in the SAS Data Analysis threads, we generated a filtered EPIC event list, which was then utilized to produce scientific products such as light curves, spectra, and other relevant data. 
A standard energy range of 0.3-10 keV was used in our study. Also, we created light curves in the energy range 0.3-2 keV (soft X-ray regime) and 2-10 keV (hard X-ray regime) to study the flux state of 3C 273 for these observations.

Light curves of source plus background and a background subtracted light curve were generated for all the sources, following the procedures suggested in the SAS Data Analysis Threads. From the cleaned data sets, source spectra were extracted from circular apertures centered on the source region. Extraction of the source spectrum was done carefully by setting the RA and Dec of 3C 273 and selecting the proper aperture size. The size of the source region varied from 40'' to 55''. For the background region, a circle of size 60'' was chosen away from the source. \\

The SAS task \textit{epatplot} is used to check for possible pile-up in the observations. For this, the pattern statistics were first checked for a circular region centred at the source. If pile-up was detected, an annular region was considered to reduce pile-up. The radius of the excluded inner area varied depending on the degree of pile-up. This region varied from 2.5'' to 7.5'' depending upon the degree of pileup.
\\
In the next section, we describe the methods we have utilised to analyse the variability such as excess variance, fractional variability, and power spectral density. Also, the various spectral models and hardness ratios are discussed.

\begin{figure*}
	\begin{minipage}{.4\textwidth} 
		\includegraphics[height=9cm]{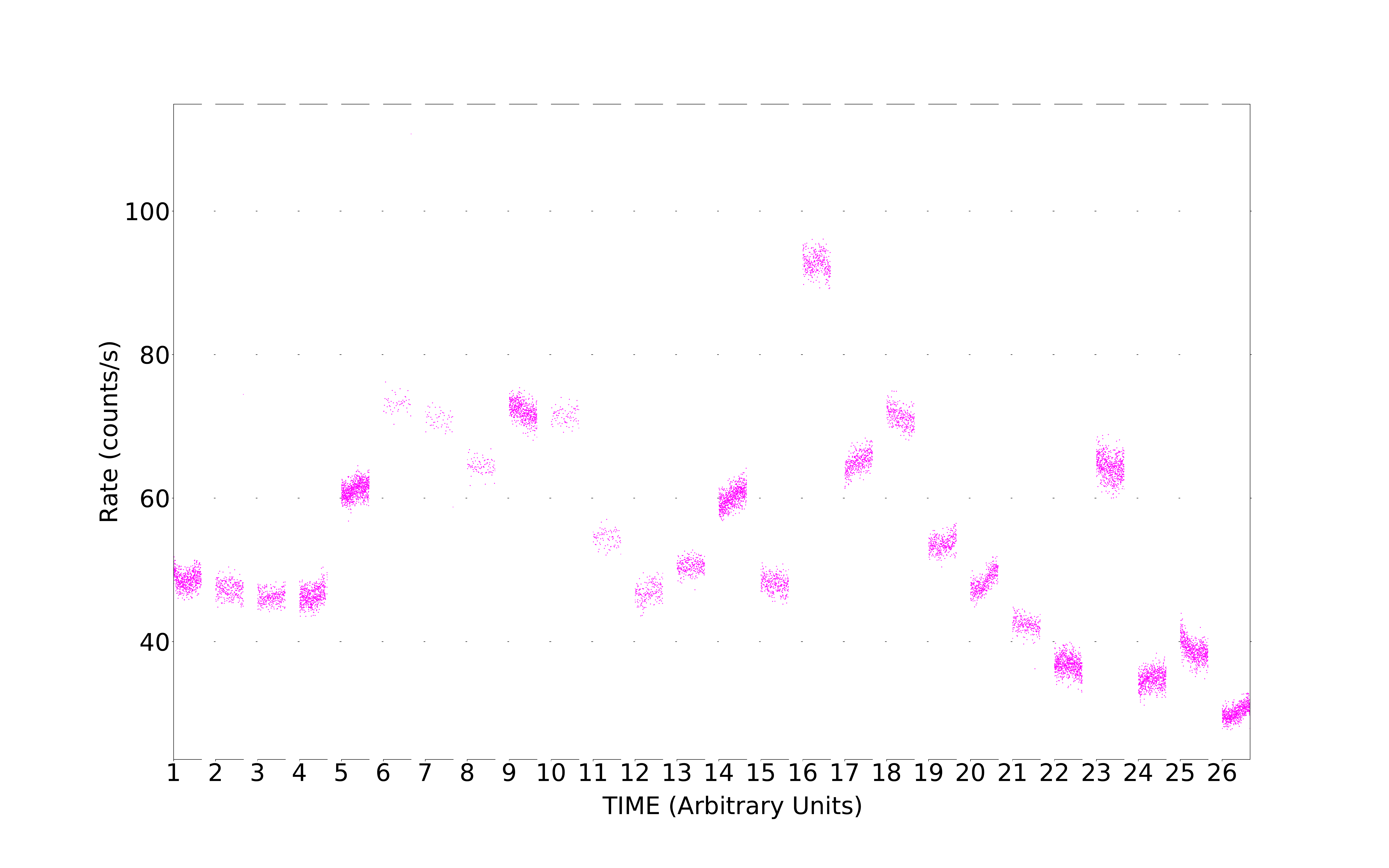}
	\end{minipage}
	\begin{minipage}{.2\textwidth}
	\  
	\end{minipage}%
	\begin{minipage}{.35\textwidth} 
		\raggedleft
		\includegraphics[height=8cm]{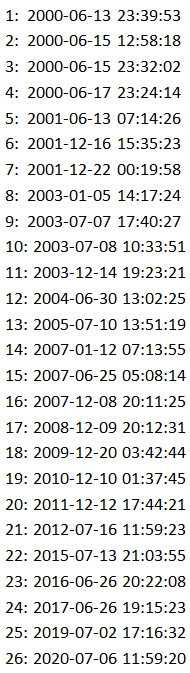}
	\end{minipage}
	\caption{X-ray \textit{XMM-Newton} observations showing light curves of 3C 273 from 2000-2020. The starting observation date and time are indicated on the right side.}
	\label{figure1}
\end{figure*}

\begin{table*}
\caption{Properties of the $\textit{XMM-Newton}$ EPIC PN observations of 3C 273 from 2000-2020. Observation ID, date of observation and length of observation are presented in Cols. 1, 2 and 3 respectively. Col. 4 and Col. 5 contain the exposure ID and mean flux in counts/s. The computed fractional variability, variability amplitude and the negative spectral power index are presented in the Cols. 6, 7 and 8 respectively.}
\label{table1}
\centering
\centering
\begin{tabular}{cccccccc}
  \hline
   Obs & Date & Exposure & Exp. ID & $\langle$F$\rangle$ & $F_{\rm var}$ & VA & $-\beta_{\rm{P}}$ \\
  & & (ks) & & (counts/s) & (\%) & &  \\\hline
   0126700301 & 2000-06-13 & 635 & S003 & $48.65\pm0.94$ & $1.209\pm 0.001$& $0.53\pm0.07$& $0.988\pm0.223$\\ 
   0126700601 & 2000-06-15 & 297 & S003 & $47.42\pm1.11$&$2.560\pm 0.126$&$1.33\pm0.48$ & \--0.078$\pm$1.987\\ 
   0126700701 & 2000-06-15 & 300 & S003  &$46.21\pm0.91$ & - & - & -\\ 
   0126700801 & 2000-06-17& 558 & S003 & $46.40\pm1.05$& $0.955\pm 0.004$ & $0.58\pm0.07$ & $1.151\pm0.183$\\ 
   0136550101 & 2001-06-13 &887 & S003 &$61.15\pm0.98$ &$0.686\pm 0.003$ &$0.54\pm0.07$ & $0.660\pm0.546$\\
   0112770101 & 2001-12-16 & 51& S001 & $73.93\pm1.39$ &
  $6.324\pm 0.429$ & $1.22\pm0.32$ & --0.729$\pm$1.350\\
   0112770201 & 2001-12-22 &51 & S001 &$70.74\pm1.22$ &
  $1.568\pm 0.254$ & $0.75\pm0.33$ & 1.981$\pm$8.279\\
   0136550501 & 2003-01-05 & 86 & S003 &$64.46\pm1.03$ & - & -& - \\
   0159960101 & 2003-07-07  & 582 & S005 & $72.13\pm1.07$& $0.731\pm 0.005$ & $0.48\pm0.07$ & $1.355\pm0.355$\\
   0112770501 & 2003-07-08  & 82 &S001 &$71.48\pm1.10$ & - & -& -\\ 
   0112771101 & 2003-12-14  & 85 & S001&$54.42\pm0.93$&$0.705 \pm0.003$ & $0.45\pm0.06$ & --2.403$\pm$2.694\\
   0136550801 & 2004-06-30 & 197& S003&$46.88\pm1.01$ &$1.104\pm 0.004$ & $0.55\pm0.08$ & 0.034$\pm$8.041\\
   0136551001 & 2005-07-10 & 277& S003&$50.52\pm0.89$ & - & -& -\\
   0414190101 & 2007-01-12  & 750 & S003&$60.12\pm1.03$ &$1.203\pm 0.001$ & $0.52\pm0.06$ & $1.516\pm0.282$\\
   0414190301 & 2007-06-25 & 321 &S003 &$48.10\pm1.06$ & $0.136\pm 0.002$ & $0.52\pm0.08$& $1.841\pm0.475$\\
   0414190401 & 2007-12-08 & 355 & S003& $92.75\pm1.21$& $0.485\pm 0.003$ & $0.40\pm0.05$& $1.875\pm0.310$\\
   0414190501 & 2008-12-09 & 404 & S003&$65.06\pm1.01$ &$1.122\pm 0.001$ &$0.49\pm0.06$ & $1.734\pm0.401$\\
   0414190601 & 2009-12-20 & 315 & S003&$71.29\pm1.06$ & $0.868\pm 0.001$ & $0.46\pm0.06$& 2.175$\pm$1.238\\
   0414190701 & 2010-12-10 & 360& S003&$53.59\pm0.92$& $0.806\pm 0.001$ & $0.51\pm0.06$& $1.716\pm0.691$\\
   0414190801 & 2011-12-12 &429 & S003&$48.28\pm0.87$ & $2.14\pm 0.002$ & $0.58\pm0.06$ & 2.153$\pm$1.007\\
   0414191001 & 2012-07-16 & 256 & S003&$42.34\pm0.91$ & - & -& -\\
   0414191101 & 2015-07-13 & 709 & S003 & $36.77\pm1.07$ & $1.022\pm 0.002$ &$0.68\pm0.09$ & $1.673\pm0.488$\\
   0414191201 & 2016-06-26  & 657  & S003 &$64.29\pm1.42$ & $0.850\pm 0.001$ & $0.56\pm0.07$ & $0.857\pm0.612$\\
   0414191301 & 2017-06-26  &655  & S003 &$34.77\pm1.05$ & $0.895\pm 0.002$ & $0.72\pm0.09$ & $1.823\pm0.690$\\
   0810820101 & 2019-07-02 & 676 & S003  &$38.85\pm1.11$ & $1.939\pm 0.003$ & $0.77\pm0.08$& $2.320\pm0.140$\\
   0810821501 & 2020-07-06 & 680  & S003 & $30.14\pm0.74$& $1.908\pm 0.002$ & $0.63\pm0.08$& $2.366\pm0.420$\\
  \hline
\end{tabular}

\end{table*}

\begin{figure*}
	\centering
	\begin{minipage}{.4\textwidth} 
		\centering 
		\includegraphics[width=.99\linewidth]{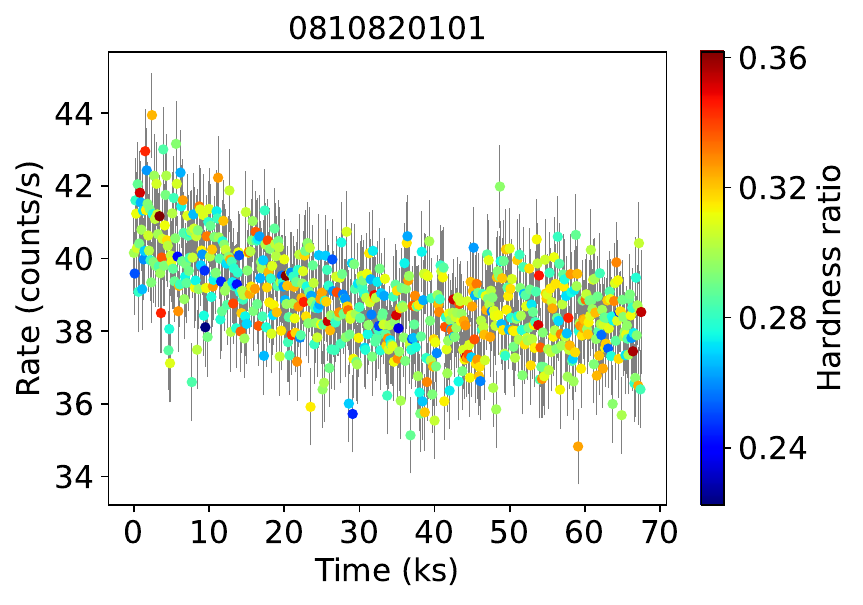}
	\end{minipage}
	\begin{minipage}{.1\textwidth} 
	\  
	\end{minipage}%
	\begin{minipage}{.4\textwidth} 
		\centering 
		\includegraphics[width=.99\linewidth]{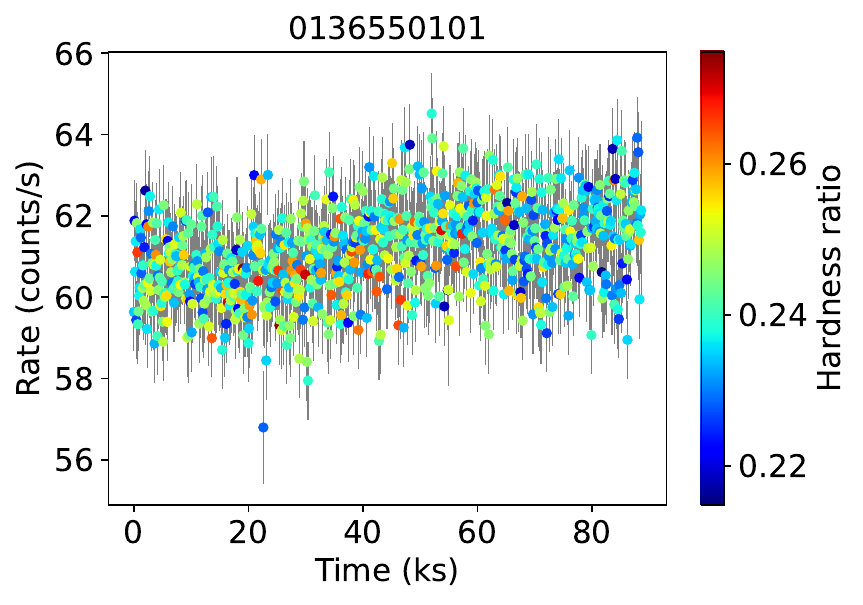}
	\end{minipage}

	\caption{Light curves of two of the observation IDs of 3C 273 (0810820101, 0136550101).}
	\label{figure2}
	\end{figure*} 

\section{Analysis and Results} \label{sec:4}

In this work, we studied all the \textit{XMM-Newton} observations of the blazar 3C 273 that were observed within the chosen period and had observation duration longer than 10 ks. The observation IDs and observation dates of the X-ray observations
are listed in Table \ref{table1}. In Figure \ref{figure1}, we present all the individual light curves of 3C 273 arranged in chronological order. The figure displays a dramatic change in the X-ray flux of the source over the past two decades. Furthermore, Figure \ref{figure2} shows the light curves of two of the individual observations with observation ID 0810820101 and 0136550101. In both observations, flux variability in the intraday timescale can be clearly observed.

In order to investigate hard X-ray variability properties of the sample sources, we performed several methods of timing and spectral analyses, which
are discussed below (see also Table \ref{table1}). The analysis methods can broadly be classified into timing and spectral analyses. For the timing analysis, we estimate excess variance, fractional variability, power spectral density (PSD) and flux distribution. Under spectral analysis, we estimate the hardness ratio and fit the spectra using various spectral models. 

\subsection{Timing Analysis}
\subsubsection{Flux-rms Relation}
In variable astrophysical sources, the nature of variability properties can be well constrained by studying any possible relationship between the root-mean-squared (rms) amplitude of variability at shorter timescale and the mean flux level at longer timescale, often known as flux-rms relation. For accreting compact objects, the flux variability has been shown to exhibit flux-rms relation \citep{uttley2001flux, uttley2005non}. A linear rms-flux relation can imply a significant correlation between the variability properties of an AGN across multiple flux states. Such a correlation could potentially indicate the presence of nonlinear processes underlying the observed variability, resulting in a flux distribution skewed towards higher flux levels. Additionally, the presence of a linear rms-flux relationship in blazar flux could suggest the involvement of multiplicative processes, such as a multiplicative coupling of disk/jet perturbations. The rms is defined as the Poisson noise corrected excess variance:
\begin{equation}\label{eq1}
\sigma^2_{\rm{XS}}=S^2-\langle{\sigma^2_{\rm{err}}}\rangle
\end{equation}
where $S^{2}= ( N-1 )^{-1} \sum_{i=1}^{N}  (\langle x \rangle -x_{\rm{i}})^{2} $ and $\langle\sigma^{2}_{\rm{err}} \rangle$ is given by:
\begin{equation}\label{eq4}
\langle\sigma^{2}_{\rm{err}}\rangle = \frac{1}{N}\sum_{i=1}^{N} \langle\sigma^{2}_{\rm{err,i}} \rangle
\end{equation}
The excess variance can be obtained by subtracting the mean squared error from the sample variance.
For determining the flux-rms relation, a light curve is split into several segments of data points, and the mean of the flux and the excess variance $S^2$ are computed for each segment. 
The uncertainty in the flux measurements can be used to estimate the uncertainty in the excess variance using an error propagation formula that can be expressed as
\begin{equation}\label{eq2}
err(\sigma_{\rm{XS}})=\frac{1}{\sigma^2_{\rm{XS}}}\sqrt{\frac{2S^2}{N-1}-\frac{Var(\sigma^2_{\rm{XS}})}{N}}.
\end{equation}

Our calculated excess variance for some segments has negative values. This is caused by the domination of  measurement errors compared to the aperiodic, frequency-dependent emission variability, commonly referred to as red noise \citep{1978ComAp...7..103P} -- which results in light curves that display more pronounced variability amplitudes on longer timescales compared to shorter ones. The variance estimates are rejected in such instances. For robust statistics, the mean fluxes and excess variances were computed for each individual observation. To examine the possible relation between them, a scatter plot of excess variance vs. mean flux is plotted which is presented in Figure \ref{figure4}, which, as it appears, does not reveal any obvious rms-flux trend.

\subsubsection{Fractional Variability and Variability Amplitude}
Fractional Variability is used to quantify the extent of variability present in light curves, and it is widely applied to test whether the processes driving the observed variability are of stationary nature \citep{vaughan2003characterizing}. The quantity combines the variance of the data points, considering the extra variance due to the uncertainties in the measurements \citep{schleicher2019fractional}. Fractional variability is given by
\begin{equation}\label{eq3}
F_{\rm var} = \sqrt{\frac{S^{2}-\langle\sigma^{2}_{\rm err} \rangle}{\langle x \rangle^{2}}}\end{equation}

The error in fractional variability is
\begin{equation}\label{eq5}
\sigma_{F_{\rm var}}=\sqrt{F^{2}_{\rm var}+\sqrt{\frac{2\langle\sigma^{2}_{\rm err} \rangle^{2}}{N\langle F \rangle^{4}}+\frac{4\langle\sigma^{2}_{\rm err} \rangle}{N\langle F \rangle^{2}}F^{2}_{\rm var}}}-F_{\rm var}
\end{equation}
\citep[see][]{bhatta2018microvariability}.
Similarly, one can estimate the observed variability by a parameter called variability amplitude (VA) used to calculate the variation in either fluxes or count rates from peak to peak. 
\begin{equation}\label{eq6}
    VA = \dfrac{F_{\rm{max}}-F_{\rm{min}}}{F_{\rm{min}}},
\end{equation}
where $F_{\rm{max}}$ and $F_{\rm{min}}$ are the maximum and minimum count rates, respectively. The uncertainty in VA ($\sigma_{\rm{VA}}$) is estimated as,

\begin{equation}\label{eq7}
    \sigma_{\rm{VA}} = (VA+1)\cdot\sqrt{\left(\dfrac{\sigma_{F_{\rm{max}}}}{F_{\rm{max}}}\right)^2+\left(\dfrac{\sigma_{F_{\rm{min}}}}{F_{\rm{min}}}\right)^2}.
\end{equation}

Fractional variability and variability amplitude measurements of blazar 3C 273 of all the observations are listed in the 6th and 7th column of Table \ref{table1}, respectively. The observations with no entries in fractional variability have negative excess variance, which suggests that the variability for those observations is not intrinsic. As Table \ref{table1} shows, the highest value of fractional variability observed is 6.3$\%$, during the year 2001 (observation ID:0112770101), whereas the lowest value (0.485$\%$) is observed during 2007 (ID 0414190401). Variability amplitude is the highest for ID 0126700601 and the lowest for ID 0112770501. However, for ID 0112770501, fractional variability is negative. The observation with the lowest non-negative value of fractional variability (0.485 $\pm$ 0.003) and the lowest variability amplitude has ID 0414190401. The mean of the fractional variability throughout the observation period is 0.019 and the mean flux is 55.413 counts/s, while the mean of variability amplitude was found to be 0.633. As suggested by the fractional variability and variability amplitude, the source is found to be moderately variable in intraday timescales throughout the observation period. This result is in line with the conclusion of \citet{pavana2022} where they have shown that light curves of 3C 273 do not exhibit large amplitude variations. 
\begin{figure}
      \includegraphics[width=1.\linewidth]{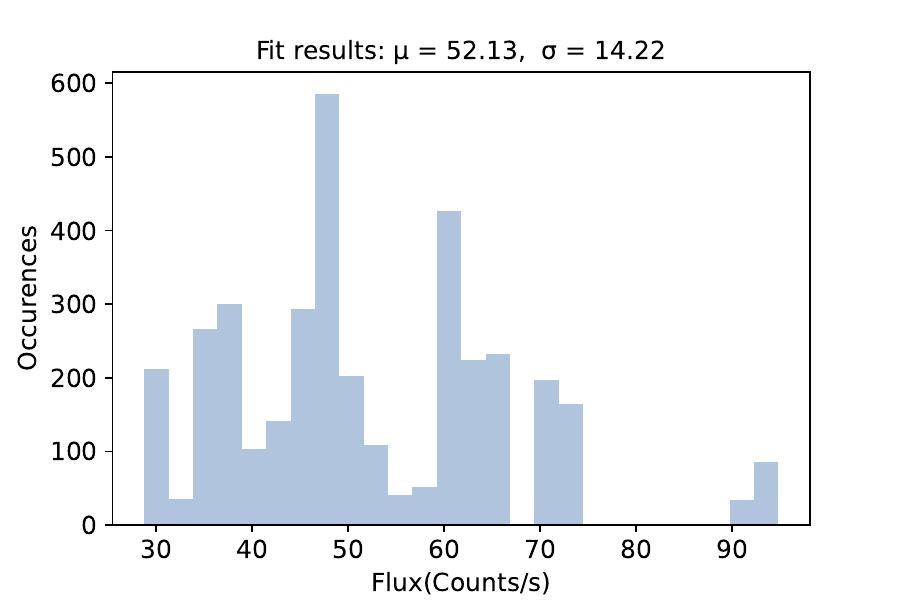}
    \caption{Flux histogram of 3C 273 using $\textit{XMM-Newton}$ PN observations during 2000-2020.}
    \label{figure3}
\end{figure}

\begin{figure}
      \includegraphics[width=8.74truecm]{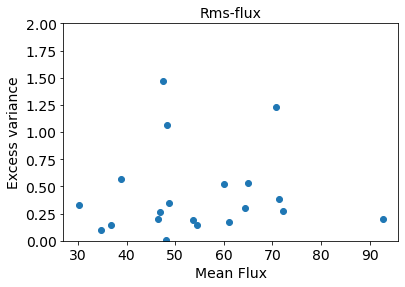}
    \caption{Flux-rms relation of the blazar 3C 273 using $\textit{XMM-Newton}$ PN observations during 2000-2020.}
    \label{figure4}
\end{figure}

\begin{figure}
    \includegraphics[width=0.99\linewidth]{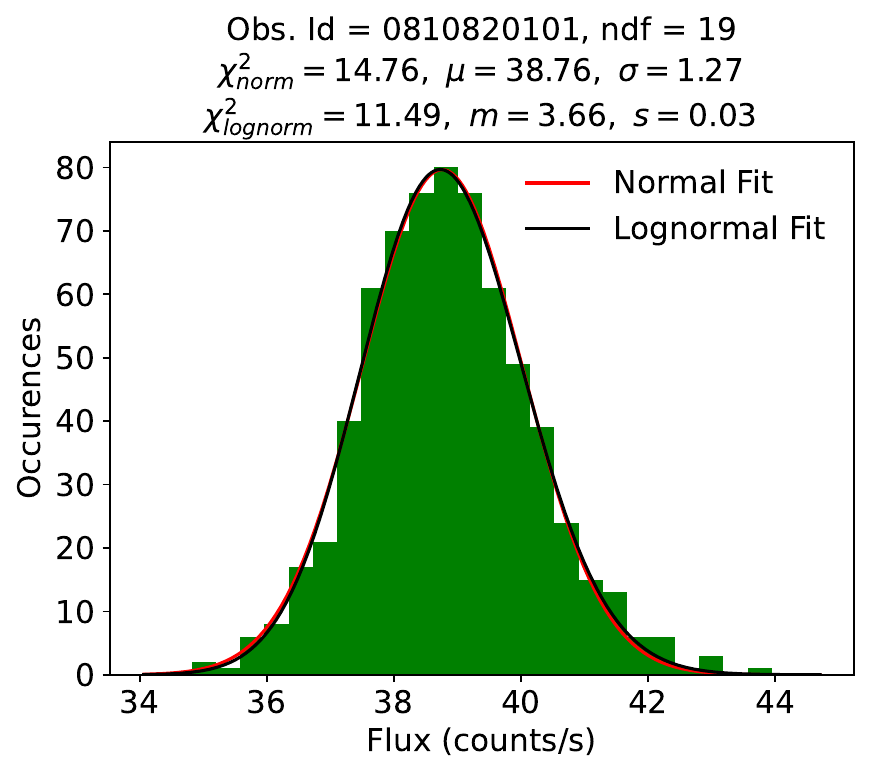}
    \caption{Flux histogram distribution of one of the observation ID of 3C 273 (0810820101).}
    \label{figure5}
\end{figure}

\begin{figure}[!ht]
    \includegraphics[width=0.99\linewidth]{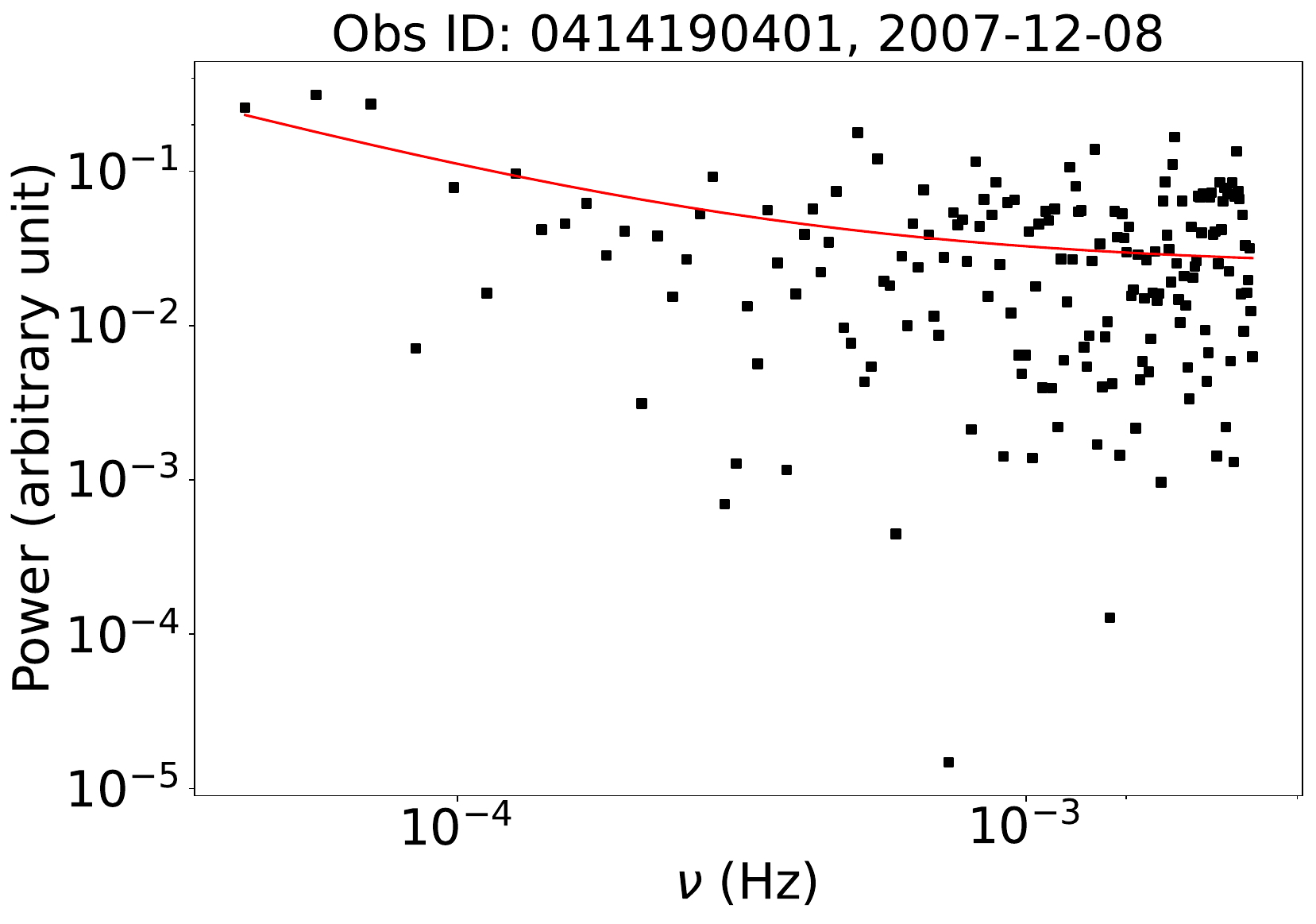}
    \caption{X-ray DFP of one of the observation IDs (0414190401) of the blazar 3C 273 on intraday timescales.}
    \label{figure6}
\end{figure}
\begin{table*}

\caption{Normal and log-normal distribution fit statistics for the X-ray flux distributions of each observation.
The mean ($\mu$) and the standard deviation ($\sigma$) of the normal fit are shown in columns 2 and 3 while the mean location (m) and the scale parameters (s) of the log-normal fit are listed in the columns 5 and 6 respectively.}
\label{table2}
\centering
\centering
\begin{tabular}{llll|llll}
 \hline
 & &  Normal fit & &  & Log-normal fit & &\\
   Obs & $\mu$ & $\sigma$ & $\chi^{2}/{\rm dof}$ & $m$ & $s$ & $\chi^{2}/{\rm dof}$ \\
  & (ks) & & (counts/s) & (\%) & & & \\
  \hline
0112770101 & 73.15 & 0.78 & 7.76/7 & 4.29 & 0.01 & 7.78/7\\
0112770201 & 71.02 & 1.31 & 4.54/3 & 4.26 & 0.02 & 4.52/3 \\
0112770501 & 71.48 & 1.08 & 3.73/4 & 4.27 & 0.02 & 3.79/4 \\
0112771101 & 54.54 & 0.98 & 11.26/4 & 4.00 & 0.02 & 11.91/4 \\
0126700301 & 48.63 & 1.16 & 7.72/14 & 3.88 & 0.02 & 8.98/14 \\
0126700601 & 47.39 & 1.15& 28.49/10 & 3.86 & 0.02 & 31.92/10 \\
0126700701 & 46.17 & 0.85 & 16.50/9 & 3.83 & 0.02 & 16.35/9 \\
0126700801 & 46.36 & 1.18 & 14.93/14 & 3.84 & 0.03 & 14.66/14 \\
0136550101 & 61.16 & 1.15 & 33.30/18 & 4.11 & 0.02 & 35.49/18 \\
0136550501 & 64.42 & 0.82 & 6.78/5 & 2.25 & 0.02 & 86.00/5 \\
0136550801 & 46.87 & 1.15 & 17.24/9 & 3.85 & 0.02 & 16.63/9 \\
0136551001 & 50.53 & 0.87 & 9.20/11 & 3.92 & 0.02 & 9.11/11 \\
0159960101 & 72.14 & 1.25 & 23.77/15 & 4.28 & 0.02 & 25.57/15 \\
0414190101 & 60.11 & 1.33 & 15.78/15 & 4.10 & 0.02 & 16.21/15 \\
0414190301 & 48.13 & 1.10 & 9.85/10 & 3.87 & 0.02 & 10.74/10 \\
0414190401 & 92.67 & 1.30 & 16.46/10 & 4.53 & 0.01 & 15.85/10 \\
0414190501 & 65.06 & 1.22 & 12.62/13 & 4.18 & 0.02 & 12.34/13 \\
0414190601 & 71.28 & 1.27 & 12.52/10 & 4.27 & 0.02 & 12.33/10 \\
0414190701 & 53.55 & 1.00 & 18.59/12 & 3.98 & 0.02 & 18.51/12 \\
0414190801 & 48.12 & 1.45 & 35.75/10 & 3.87 & 0.03 & 30.43/10 \\
0414191001 & 42.35 & 0.84 & 12.13/12 & 3.75 & 0.02 & 11.93/12 \\
0414191101 & 36.76 & 1.10 & 16.46/18 & 3.60 & 0.03 & 15.52/18 \\
0414191201 & 64.32 & 1.51 & 18.47/16 & 4.16 & 0.02 & 20.08/16 \\
0414191301 & 34.81 & 1.12 & 21.70/18 & 3.55 & 0.03 & 26.93/18 \\
0810820101 & 38.76 & 1.27 & 14.76/19 & 3.66 & 0.03 & 11.49/19 \\
0810821501 & 30.07 & 0.95 & 18.62/13 & 3.40 & 0.03 & 16.91/13 \\\hline
\end{tabular}

\end{table*}


\subsubsection{Flux distributions}
A proper characterization of blazar flux distribution can be quite useful in understanding the nature of physical processes contributing to the emission mechanism, and the origin and nature of their variability. With this goal, we constructed histograms of the observed flux and fitted two different probability density functions (PDFs), a normal distribution and a log-normal distribution, to the flux histograms. A normal distribution is defined by
\begin{equation}
    N_{\rm{norm}}(x) = \dfrac{1}{\sqrt{2\pi}\sigma}\exp(-\dfrac{(x-\mu)^2}{2\sigma^2}),
\end{equation}
where $\mu$ and $\sigma$ are the mean and the standard deviation expressed in the units of flux ($\text{counts}/\text{cm}^2/\text{s}$) respectively. A clear evidence of normality of the flux distribution is indicative of the fact that the observed flux originates from additive processes, i.e., the observed flux is the sum of fluxes contributed by many processes possibly occurring in different regions. On the other hand, a log-normal distribution is defined by
\begin{equation}
    N_{\rm{lognorm}}(x) = \dfrac{1}{\sqrt{2\pi}sx}\exp(-\dfrac{(\ln x-m)^2}{2s^2}),
\end{equation}
where $m$ (expressed in the units of the natural logarithm of flux) and $s$ are the mean locations and the scale parameters of the distribution, respectively. A flux distribution characterized by such PDF, in contrast, is indicative of the observed flux being an outcome of multiplicative processes undergoing in blazars \citep{uttley2005non}.

A flux histogram of the source, obtained using the \textit{XMM-Newton} PN observations made during various observation periods from 2000-2020, is shown in Figure \ref{figure3}. The distribution of the flux is clearly neither normal nor log-normal but features several peaks (modes). Such multi-modal nature of the flux histogram could be the consequence of multiple emission or activity states. In order to delve further, we obtained flux histograms for each observation and fitted the normal and log-normal distributions to the histograms (shown in Figure \ref{figure5}). The fit results are tabulated in Table \ref{table2}. It is interesting to note that on an observation-by-observation basis, there are no clear multi-modal flux distributions. For most of the observations, both the normal and the log-normal distributions describe the flux histograms alike. Based on the reduced chi-squared values from the fit, the source histogram can equally be represented by both of the considered two distributions.

\subsubsection{Power spectral density analysis}
The statistical properties of the observed variability can be investigated using PSD analysis. The PSD shape in the frequency domain can be measured by employing a Discrete Fourier periodogram (DFP).
It is a measure of the variability power at a given temporal frequency (or at a given timescale). For a time series sampled at times $t_j$ with $j = 1, 2, ..., n$, DFP for a given temporal frequency $\nu$ can be given by,
\begin{equation}\label{eq8}
    P(\nu) = \dfrac{T}{\langle x\rangle^2N^2}\left|\sum_{j=1}^n x(t_j)e^{-i2\pi\nu t_j}\right|^2,
\end{equation}
where $T$ and $\langle x\rangle$ are the total observation duration and the mean flux of the time series, respectively. In Equation \ref{eq8}, with the given normalization, the periodogram is expressed in the units of $\rm (rms/mean)^2 Hz^{-1}$.\\ 
Blazar 3C 273 is moderately bright in the X-ray band such that the light curves presented here are mostly evenly spaced, having a time bin of 100 s. Nonetheless, some light curves are not completely evenly spaced. In such a case, we filled such gaps using linear interpolation.
The periodograms computed in such a way usually exhibit scatter around the true underlying PSD. These periodograms were fitted \textit{PL} PSD model, 
\begin{equation}
    P(\nu)=N_0\nu^{-\beta_{\rm{P}}}+C,
\end{equation}
where N$_0$, $\beta_{\rm{P}}$ and $C$ represent a normalization factor, spectral power index and the Poisson noise level, respectively \citep{Bhatta2016}. As a representative case, the source periodogram for the source ID 0414190401 and the fitted \textit{PL} model are shown by black symbols and red curve, respectively, in Figure \ref{figure6}. The model PSD were fitted using the python package \textit{SciPy} \footnote{\url{https://docs.scipy.org/doc/scipy/reference/generated/scipy.optimize.curve_fit.html}}. It should be noted that the uncertainties in the PSD index reflect the quality of fitting, suggesting the larger uncertainties for poor fitting to the observations. The best-fit slope indexes of the PSD models are listed in the 8th column of Table \ref{table1}. For the observation IDs with null F$_{\rm var}$, the PSD slopes are not presented in the table, as the variability observed is less than the measured uncertainties. This means the light curve is mainly dominated by Poisson noise and the PSD slopes are close to zero. It is seen that the PSD slopes range between $-2.403$ to $2.366$ with a caveat that some slopes have larger uncertainty. We note that in a similar work by \citet{pavana2022} the \textit{PL} slopes of the PSD range between $ 1.7$ to $2.7$. It is pointed out that the relatively flatter PSD slopes are consistent with the smaller values for fractional variability (see column 6 of Table~\ref{table1}), and are indicative of the fact that the sources are only moderately variable on intraday timescales. But in the case of highly variable blazars, e.g. S5 0716+714, the slopes can be steeper \citep[see the similar recent work by][]{mohorian2022}.

\begin{figure}
    \includegraphics[width=.99\linewidth]{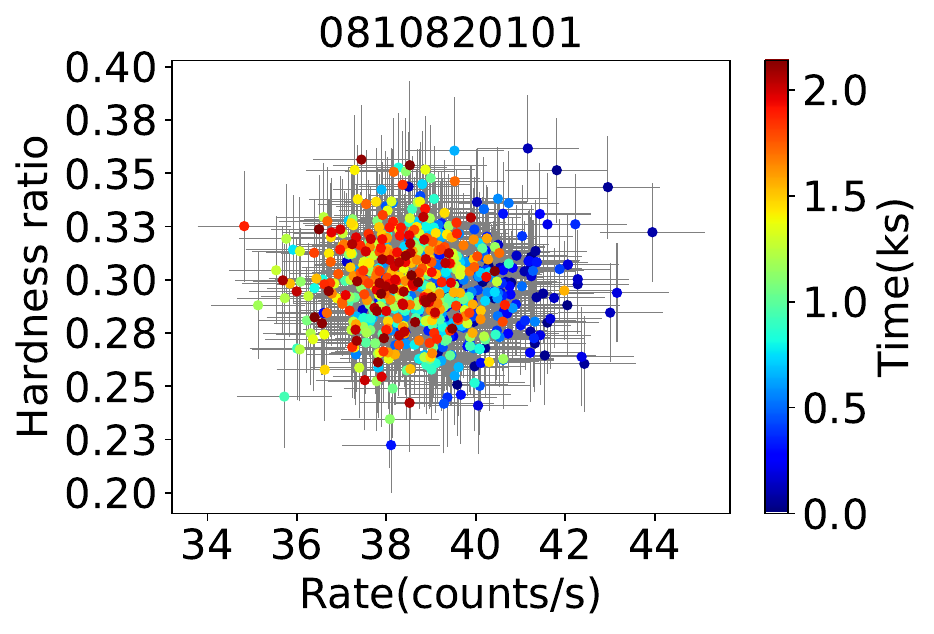}
    \caption{Hardness ratio of one of the observation IDs of 3C 273 (0810820101).} 
    \label{figure7}
\end{figure}

\begin{figure}
    \includegraphics[height=.99\linewidth, angle=-90]{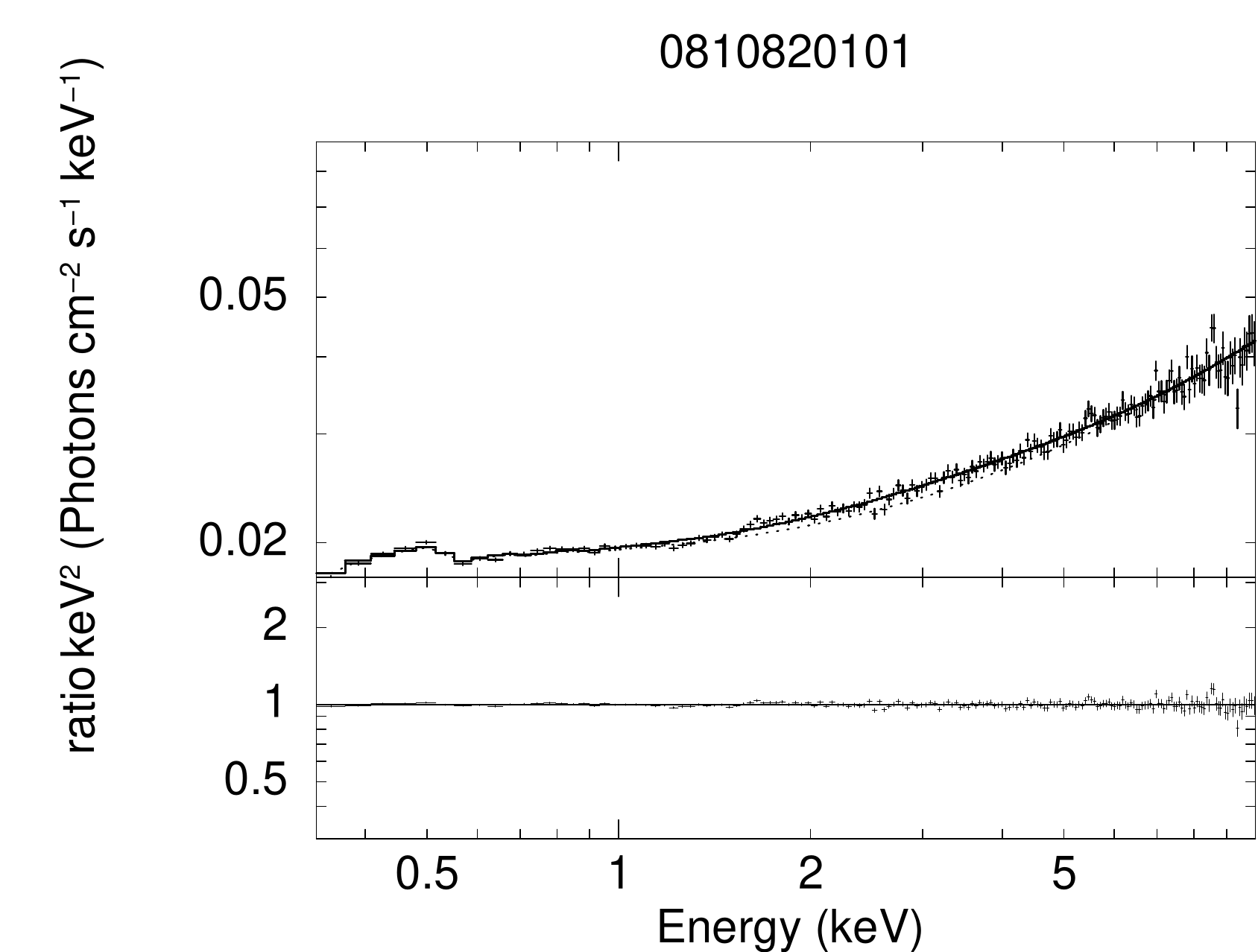}
    \caption{Spectral fitting of log-parabola model to one of the observation IDs of 3C 273 (0810820101).}
    \label{figure8}
\end{figure}
\subsection{Spectral Analysis}
In this section, we describe the various techniques that were used to examine the spectral properties of the blazars in their diverse flux states.
\subsubsection{Hardness Ratio}
To study spectral variations over time, we used the hardness ratio analysis.
We define hardness ratio (HR) as:
\begin{equation}\label{eq9}
    HR = \frac{H}{S},
\end{equation}
where \textit{H} is the flux in the hard X-ray energy band (2-10 keV) and \textit{S} is the flux in the soft energy band (0.3-2 keV). We have chosen this definition to constrain the values between 0 and 1 due to the typical AGN spectral shape, which usually declines towards hard X-ray energies. The error in hardness ratio is given as:
\begin{equation}\label{eq10}
    \sigma_{HR} = HR\cdot\sqrt{\left(\dfrac{\sigma_{F_{\rm{hard}}}}{F_{\rm{hard}}}\right)^2+\left(\dfrac{\sigma_{F_{\rm{soft}}}}{F_{\rm{soft}}}\right)^2}.
\end{equation}

A variation in the hardness ratio exhibits spectral variability during the observations. It is also useful to test the possible correlation/inverse correlation with the source flux state (count rate) in order to establish the nature of flux variations.
We tested for possible correlation between the HR and the flux state by plotting these quantities against each other, and one representative plot is shown in Fig. \ref{figure7}. We did not find any notable correlation in almost all cases. This finding is similar to the result of \citep{pavana2022} where the authors did not find significant changes in HR with time. 
Moreover, we could not detect any signs of a hysteresis loop in the HR-count rate plots.

\subsubsection{Spectral Modeling}
A study based on spectral modelling of all 26 XMM-Newton observations of the blazar 3C 273 was carried out following the standard procedure of spectral analysis of X-ray observations. 
During the data processing, the source spectrum, response matrix file (RMF), ancillary response file (ARF), and background spectrum were created for each observation and linked together. The spectrum was then binned in order to have 50 counts in each bin for all observations.
This grouped spectrum file was then analysed using the X-ray spectral fitting package $\textit{XSPEC}$ version 12.11.0 \citep{xspec1996} provided by High Energy Astrophysics Science Archival Research Center (HEASARC),
NASA/GSFC\footnote{https://heasarc.gsfc.nasa.gov/}. Various models in $\textit{XSPEC}$ were used for the spectral fitting. Reduced $\chi$2 value was used to determine the best model. The spectral fit was done in the energy range of 0.3-10 keV.
In all the spectral fitting, we have used the $\textit {tbabs}$ model, which considers X-ray absorption due to Galactic neutral hydrogen towards the observer.
The input parameter, $N_{\rm H}$ in this model, is the equivalent hydrogen column (in units of $10^{22}$ atoms cm$^{-2}$).
We fixed the neutral H column density value at $N_{\rm H}=1.69\times10^{20}$cm$^{-2}$ following \citet{dickey1990hi}.

We first fit the spectra with a simple \textit{PL} absorbed by the
Galactic hydrogen column along the line of sight to 3C 273.
The expression for \textit{PL} model is:
\begin{equation} \label{eq14}
    A(E) = K E^{-\alpha},
\end{equation}
where $\alpha$ is the first parameter, which is known as the photon index (dimensionless). The next parameter is norm $K$, photons/keV/cm$^2$/s at 1 keV.  The spectrum of the non-thermal emission resulting from a power-law injection of high-energy particles in the jet magnetic field often mimics the particle distribution. In particular, the index of the synchrotron emission spectrum ($\alpha$) is related to the power-law index of the particle distribution ($p$) by the equation $\alpha$ = ($p$ - 1)/2.
Another model that we used is the $\textit{log-parabola}$ model (\textit{LP}; \citealt{2004A&A...413..489M,2006A&A...448..861M}). The corresponding expression is:

\begin{equation} \label{eq15}
A(E) = K(E/E_{\rm{pivot}})^{-a-b\log(E/ E_{\rm{pivot}})},
\end{equation}
where $E_{\rm pivot}$ stands for fixed pivot energy, $a$ is the slope at the pivot energy and $b$ is the curvature term. It was observed that fitting
\textit{log-parabola} model to the source X-ray spectra yielded reduced $\chi^{2}$ values that were smaller than the ones we obtained when fitting the spectra using \textit{PL} model.  The $LP$ spectral model describes the spectrum resulting from a particle acceleration scenario, where a statistical acceleration mechanism relies on energy-dependent probabilities. As the energy of a relativistic particle increases, the probability of further energy gain through particle acceleration decreases. To further investigate the source X-ray emission for possible contribution from the accretion disk, the spectra were also fitted with a \textit{black-body} (\textit{BB}) model added to the \textit{LP} model. This indeed resulted in the reduced $\chi^{2}$ values that were closer to unity in most of the observations, indicating the combined model provided a better description of the source emission. The \textit{BB} model can be expressed as
\begin{equation}\label{eq16}
    \frac{dN}{dE} = \frac{N_0\cdot E^2}{(kT)^4\cdot(\exp(E/kT)-1)},
\end{equation}
where $kT$ is temperature in keV and $N_0$ is the normalization parameter. The results were further verified using F-test. The results are tabulated in the 8th column of Table \ref{table3}.

Finally, to further examine the source spectra for the possible presence of \emph{break energy}, the \textit{broken power-law} (\textit{BPL}) model was used for the spectral analysis. The \textit{BPL} model consists of two \textit{power-law} indexes glued at break energy.
The expression for \textit{BPL} model is:
\begin{equation}\label{eq17}
A(E)=\left\{\begin{array}{ll}
K E^{-\alpha_{1}}, & \text { if } E \leq E_{\rm break}, \\
K E_{\rm break}^{\alpha_{2}-\alpha_{1}}(E / 1 k e V)^{-\alpha_{2}}, & \text { if } E>E_{\rm break},
\end{array}\right.
\end{equation}
where $E_{\rm break}$ is the break point for the energy in keV, $\alpha_{1}$ is the \textit{PL} photon index for $E \leq E_{\rm break}$, $\alpha_{2}$ is the \textit{PL} photon index for $E \geq E_{\rm break}$. A broken power-law X-ray spectrum indicates the possible presence of high energy relativistic electrons which rapidly cool  by radiation resulting in a break in the emission spectrum. 

In the beginning, a simple \textit{PL} model is used to fit all the observations considered, and we found that it does not provide a good fit (the reduced $\chi^2$ is very high in all cases as evident from Table~\ref{table3}). In the next step, we changed our base model to the \textit{LP} model and found that the fit improves significantly as compared to the \textit{PL} model. However, the fit still is not the best one for many of the observations. This result hints at the requirement for an additional component in the spectral fitting.
It is clear from the fitting parameters presented in Table~\ref{table3} that, for most of the observations, except for 0112770101 and 0112770501, \textit{LP} added with a \textit{BB} model is a better fit. For the mentioned observation IDs, \textit{BPL} was a better fit, but by only a small margin. In Fig.~\ref{figure8}, we show the spectrum of one of the observations fitted with a \textit{log-parabola} model. The temperature as estimated by the \textit{log-parabola+black-body} model is in the range 0.75-1 keV. We found that the slope of the best-fit model, $\alpha$ varies between 1.8-2.1 among the observations. 
\section{Discussion}\label{sec:5}
\subsection{Variability Properties}
A study of several episodes of intraday flux and spectral variability spanning over nearly two decades reveals many interesting variability features that help constrain the underlying physics of the AGN. As seen in Figure \ref{figure1}, over the period the source displays a flux change of nearly three times with a long-term fractional variability of 27.2\% in the X-ray band. It is noted that the FV in the X-ray regime is larger than the 14\% FV in the optical band as reported by \citet[][]{Bhatta2021} and considerably smaller than the 94\% FV in the Fermi/LAT $\gamma$-ray band as reported by \citet[][]{bhatta2020nature}, although the latter two works include the observations from the last 12 years only. 

\citet{bhattacharyya2020blazar} estimated the fractional variability of 3C 273 to be less than 1 percent for \textit{XMM-Newton} observations during 2001 and 2007 which is consistent with our results. Very recently, \citet{pavana2022} analysed 23 light curves spanning 20 years of \textit{XMM-Newton} observations and found that nine light curves showed small (0.71 per cent to 3.04 per cent) amplitude variations while 14 light curves did not show any significant variability in X-ray energy band.

The observed variability in blazars can largely be attributed to the processes taking place in the accretion disk and jets of AGN. In the scenario involving the accretion disk, the variability arises from magnetohydrodynamic instabilities caused by fluctuations in the accretion rates, viscosity parameter, and magnetic field.  Similarly, in the scenarios involving the jets, several variability models have been widely discussed. These models mainly involve particle acceleration via shock waves, magnetic re-connection, and turbulence  in the jets \citep[see e. g.][]{2009MNRAS.395L..29G,1998A&A...333..452K,maraschi1992jet,marscher1985models}.  The X-ray variability observed in the timescale of tens of years can be the result of combined mechanisms. However, the intra-night variability can arise due to the rapid cooling of relativistic electrons via dissipative processes such as synchrotron and inverse-Compton emission.

\subsection{Flux distribution}
The study of flux distribution helps us in determining whether the variability is arising from multiplicative or additive processes. For the overall flux distribution throughout the observation period (2000-2020), we found that the distribution is neither normal nor log-normal. However, the distribution of the individual observations is described by a normal distribution, and in some cases by a log-normal distribution.

With the objective of getting an insight into the possible processes responsible for the observed X-ray variability of 3C 273, we investigated its flux distribution. The overall flux distribution throughout the observation period (2000-2020) featured several modes and clearly did not agree with either of the two distributions we considered -- the normal and the log-normal distributions. 

\begin{longtable*}[h]{|c|c|c|c|c|c|c|c|}
\caption{Spectral properties of 3C 273. Col. 1: observation ID; Col. 2: Spectral models, \textit{power-law} (\textit{PL}), \textit{log-parabola} (\textit{LP}), \textit{log-parabola+black-body} (\textit{LP+BB}), \textit{broken power-law} (\textit{BPL}), \textit{broken power-law+black-body} (\textit{BPL+BB}); Col. 3: $\chi^2$/degrees of freedom; Col 4: photon index (\textit{PL}), curvature parameter (\textit{LP}), high-energy photon index (\textit{BPL}); Col. 5: low-energy photon index (\textit{LP}), low-energy photon index (\textit{BPL}); Col. 6: break energy in keV; Col. 7: black-body temperature (keV); Col. 8: F-test results.}
\label{table3}
\\\hline
Obs & Model & Reduced & $\alpha$/Ep/$\alpha_{1}$ & $\beta$/$\alpha_{2}$&$E_{\rm break}$ & BB temp & F-test \\
   & & chi-square & & & keV & keV & \\
\endfirsthead
\multicolumn{4}{c}%
{{\bfseries \tablename\ \thetable{} -- continued from previous page}} \\
\hline
 Obs & Model & Reduced & $\alpha$/Ep/$\alpha_{1}$ & $\beta$/$\alpha_{2}$&$E_{\rm break}$ & BB temp & F-test \\
   & & chi square & & & keV & keV & \\

\hline
\endhead

\hline \multicolumn{8}{|r|}{{Continued on next page}} \\ \hline
\endfoot

\endlastfoot

\hline
   0126700301 &PL & 9248.25/176(51.54) & $1.812\pm.001$& -& - &- & \\
    & LP &835.36/175(4.77)  & $1.901\pm.001$ & $-0.264\pm.002$&- &- & \\ 
     & LP+BB & 384.79/173(2.22) & $1.945\pm.002$ & $-0.290\pm.005$ &- &$0.994\pm.02$ & 100.702\\ 
   & BPL&769.84/174(4.42) & $2.034\pm.003$& $1.664\pm.002$& $1.148\pm.009$&- & \\
   & BPL+BB&917.34/172(5.33) & $2.029\pm.003$& $3.243\pm.070$& $2.885\pm.024$&$2.071\pm.014$ &\\
  0126700601  &PL &4284.43/171(25.05) &  $1.797\pm.001$&-&-&- &\\
     & LP &340.38/170(2.00) & $1.905\pm.002$&$-0.315\pm.004$&- &- &\\ 
     & LP+BB &209.31/168(1.24)  & $1.948\pm.004$& $-0.345\pm.008$&-&$0.969\pm.052$ & 52.288 \\ 
  & BPL &384.52/169(2.27) & $2.046\pm.005$& $1.617\pm.003$& $1.207\pm.015$& - & \\
  & BPL+BB&412.55/167(2.47) & $2.053\pm.005$& $3.362\pm.127$& $2.722\pm.036$&$2.03\pm.022$ & \\
   0126700701 &PL & 5399.63/174(31.03)&  $1.800\pm.001$&-&-&- & \\
     & LP &555.63/173(3.21) &  $1.903\pm.002$&$-0.299\pm.004$&-& - &\\
     & LP+BB &234.98/171(1.37) & $1.960\pm.003$& $-0.342\pm.007$&-&$0.956\pm.032$ & 115.99 \\
   & BPL &459.55/172(2.67) & $2.069\pm.005$& $1.639\pm.003$& $1.108\pm.012$& - &\\
   & BPL+BB&288.13/170(1.69) & $2.159\pm.007$& $0.958\pm.024$& $3.289\pm.068$&$0.942\pm.015$ & \\
   0126700801  &PL &8550.66/175(48.86) &  $1.799\pm.001$&-&-& - &\\
   & LP&871.08/174(5.00) &  $1.909\pm.001$&$-0.321\pm.003$&- & - &\\
    & LP+BB & 314.15/172(1.82)& $1.975\pm.003$& $-0.375\pm.006$&-&$0.929\pm.023$ & 151.58\\
   & BPL &632.54/173(3.65) & $2.099\pm.004$& $1.630\pm.002$& $1.078\pm.009$& - &\\
   & BPL+BB & 869.67/171(5.08) & $2.071\pm.004$& $3.923\pm.116$& $2.980\pm.024$&$2.014\pm.015$ &\\
  0136550101 &PL &37981.60/176(215.80) &  $1.912\pm.001$&-&-&- &\\
    & LP &2092.45/175(11.96) &  $2.030\pm.001$&$-0.391\pm.001$&-&- &\\
    & LP+BB & 512.00/173(2.95)& $2.086\pm.001$& $-0.424\pm.003$&-&$0.977\pm.015$ & 265.47\\
   & BPL &1962.30/174(11.27) & $2.220\pm.002$& $1.676\pm.001$& $1.165\pm.004$& - &\\
  & BPL+BB&4700.67/172(27.32) & $2.167\pm.001$& $3.147\pm1.00$& $14.434\pm1.00$&$1.800\pm.007$ &\\
   0112770101 &PL &1271.83/171(7.43) &  $1.843\pm.003$&-&-&- &\\
    &LP &276.75/170(1.63) &  $1.917\pm.003$&$-0.244\pm.007$ &-&- &\\
    & LP+BB &206.89/168(1.23) & $1.965\pm.006$& $-0.283\pm.013$&-&$0.877\pm.060$ & 28.195\\
   & BPL &203.54/169(1.21) & $2.046\pm.009$& $1.696\pm.006$& $1.146\pm.027$& - &\\
   & BPL+BB&256.26/167(1.53) & $2.042\pm.007$& $3.712\pm.224$& $3.080\pm.057$&$2.108\pm.041$ & \\
   0112770201 &PL &1063.19/170(6.25) &  $1.783\pm.003$&-&-&-&\\
   &  LP&259.33/169(1.53) &  $1.862\pm.004$&$-0.225\pm.007$&-& -& \\
   &  LP+BB &183.05/167(1.06) & $1.914\pm.007$& $-0.253\pm.014$&-&$1.018\pm.074$ & 34.587\\
   & BPL &194.38/168(1.15) & $1.984\pm.010$& $1.655\pm.006$& $1.142\pm.030$& -&\\
   & BPL+BB&215.99/166(1.30) & $1.985\pm.008$& $3.248\pm.193$& $2.949\pm.066$&$2.087\pm.040$ & \\
   0136550501  &PL &2621.49/166(15.79) &  $1.989\pm.002$&-&-&- &\\
     & LP &375.05/165(2.27) &  $2.073\pm.002$&$-0.308\pm.006$&- & - &\\
    &  LP+BB &228.51/163(1.40) & $2.126\pm.005$& $-0.338\pm.010$&-&$0.912\pm.046$ & 51.944\\
   & BPL &271.66/164(1.65) & $2.240\pm.007$& $1.802\pm.005$& $1.110\pm.017$& - &\\
   & BPL+BB&341.90/162(2.11) & $2.228\pm.006$& $4.345\pm.233$& $2.938\pm.041$&$1.855\pm.023$ &\\
   0159960101 &PL &5967.03/175(34.09)  &  $1.927\pm.001$&-&-&- &\\
    & LP &559.34/174(3.21) &  $1.980\pm.001$&$-0.178\pm.002$ &-& -&\\
     & LP+BB  &365.55/172(2.12) &$2.006\pm.002$ & $-0.203\pm.004$&-&$0.780\pm.029$ & 45.326\\
   & BPL &454.43/173(2.62) & $2.065\pm.027$& $1.822\pm.196$& $1.162\pm.012$& - &\\
   & BPL+BB&591.44/171(3.45) & $2.065\pm.024$& $2.935\pm.047$& $2.705\pm.022$&$2.060\pm.014$ & \\
  0112770501  &PL  &786.90/166(4.74) &  $1.933\pm.002$&-&-&- &\\
    & LP  &228.52/165(1.38) &  $1.978\pm.003$&$-0.156\pm.006$&- & -&\\
    &  LP+BB  & 210.22/163(1.28)& $2.001\pm.006$& $-0.183\pm.012$&-&$0.708\pm.079$&7.051\\
   & BPL &208.67/164(1.27) & $2.047\pm.007$& $1.836\pm.005$& $1.214\pm.039$& - &\\
   & BPL+BB&216.47/162(1.34) & $2.054\pm.006$& $2.973\pm.130$& $2.763\pm.058$&$2.129\pm.046$ & \\
   0112771101  &PL &1753.10/166(10.56) &  $1.876\pm.002$&-&-&-&\\
   &  LP  &264.84/165(1.61) &  $1.962\pm.003$&$-0.275\pm.006$&-&- &\\
    &  LP+BB  &175.56/163(1.08) & $2.013\pm.006$& $-0.316\pm.012$& &$0.882\pm.054$ &41.192\\
 & BPL &211.51/164(1.29) & $2.117\pm.009$& $1.722\pm.005$& $1.097\pm.021$& - &\\
   & BPL+BB&249.28/162(1.54) & $2.102\pm.007$& $3.758\pm.206$& $2.883\pm.048$&$1.985\pm.030$& \\
  
   0136550801 &PL &2047.29/168(12.18) &  $1.882\pm.002$&-&-&-&\\
    & LP &241.71/167(1.45) &  $1.959\pm.002$&$-0.265\pm.005$&- & - &\\
    &  LP+BB &177.89/165(1.08) &$1.994\pm.005$ & $-0.282\pm.010$& &$0.965\pm.076$ & 29.418\\
   & BPL &199.23/166(1.20) & $2.084\pm.006$& $1.721\pm.005$& $1.177\pm.021$& - &\\
   & BPL+BB&257.47/164(1.57) & $2.086\pm.006$& $3.293\pm.149$& $2.755\pm.049$&$1.993\pm.029$ &\\
   0136551001 &PL &7380.60/175(42.17)  &  $1.831\pm.001$&-&-&-&\\
    & LP &676.30/174(3.89) &  $1.934\pm.001$&$-0.321\pm.003$&- & -&\\
     &  LP+BB & 312.53/172(1.82)& $1.990\pm.003$& $-0.371\pm.006$& &$0.884\pm.026$& 99.517\\
   & BPL & 497.00/173(2.87)& $2.119\pm.005$& $1.653\pm.002$& $1.090\pm.010$& -&\\
   & BPL+BB&354.56/171(2.07) & $2.202\pm.006$& $0.883\pm.023$& $3.367\pm.063$&$0.939\pm.013$& \\
   0414190101 &PL &18561.08/176(105.46)  &  $1.589\pm.001$&-&-&-&\\
    &  LP  &1597.96/175(9.13) &  $1.713\pm.001$&$-0.299\pm.002$&- & -& \\
    & LP+BB  & 426.10 /173(2.46)&$1.783\pm.002$ & $-0.382\pm.004$& &$0.884\pm.013$&236.517\\
  & BPL &1293.54/174(7.43) & $1.891\pm.003$& $1.443\pm.001$& $1.102\pm.006$& -&\\
  & BPL+BB&604.11/172(3.51) & $2.015\pm.004$& $0.679\pm.012$& $3.196\pm.031$&$0.954\pm.007$ &\\
   0414190301 &PL &2881.61/173(16.66) &  $1.702\pm.001$&-&-&-&\\
    & LP &405.01/172(2.35) &  $1.787\pm.002$&$-0.229\pm.004$&- & - &\\
    & LP+BB &281.56 /170(1.66) &$1.830\pm.004$ & $-0.275\pm.008$&-&$0.859\pm.041$ & 37.049\\
   & BPL &407.60/171(2.38) & $1.914\pm.006$& $1.586\pm.003$& $1.107\pm.017$& - &\\
   & BPL+BB&631.81/169(3.74) & $1.860\pm.004$& $1.182\pm1.00$& $12.180\pm1.00$&$2.005\pm.030$ &\\
   0414190401 & PL &13753.50/176(78.14) &  $1.811\pm.001$&-&-&-&\\
    & LP &1172.75/175(6.70) &  $1.910\pm.001$&$-0.291\pm.002$&-&- &\\
    &  LP+BB & 282.07 /173(1.63)&$1.969\pm.002$ & $-0.345\pm.004$& - &$0.885\pm.016$ &271.558\\
   & BPL &713.17/174(4.09) & $2.079\pm.003$& $1.654\pm.001$& $1.089\pm.007$& -&\\
   & BPL+BB&1300.93/172(7.56) & $2.055\pm.002$& $3.679\pm.075$& $2.931\pm.018$&$2.038\pm.011$ &\\
   0414190501 & PL&12893.38/175(73.68) &  $1.737\pm.001$&-&-&-&\\
    & LP &914.90/174(5.26) &  $1.849\pm.001$&$-0.315\pm.002$&-&-&\\
    &  LP+BB &246.89/172(1.43) &$1.906\pm.002$ & $-0.358\pm.005$&-&$0.987\pm.023$&231.337\\
   & BPL &585.97/173(3.39) & $2.009\pm.003$& $1.559\pm.002$& $1.168\pm.008$& -&\\
   & BPL+BB&1055.34/171(6.17) & $2.002\pm.003$& $3.503\pm.081$& $2.975\pm.022$&$2.122\pm.013$ &\\
   0414190601  &PL &11753.83/176(66.78) &  $1.762\pm.001$&-&-& -&\\
    & LP &615.53/175(3.52) &  $1.867\pm.001$&$-0.321\pm.002$&-& -&\\
    &  LP+BB &283.10/173(1.64) &$1.908\pm.002$ & $-0.349\pm.005$&-&$0.991\pm.034$& 100.985\\
   & BPL &674.30/174(3.88) & $2.018\pm.003$& $1.571\pm.002$& $1.197\pm.009$& -&\\
   & BPL+BB&810.46/172(4.71) & $2.021\pm.003$& $3.562\pm.089$& $2.907\pm.023$&$2.130\pm.014$ & \\
   0414190701  &PL &9555.39/176(54.29) &  $1.782\pm.001$&-&-&-&\\
    & LP &719.13/175(4.11) &  $1.882\pm.001$&$-0.308\pm.003$&-&-&\\
    & LP+BB &278.24 /173(1.61) &$1.931\pm.002$ & &$-0.332\pm.005$&$1.057\pm.033$ &136.273\\
   & BPL &543.27/174(3.12) & $2.037\pm.003$& $1.599\pm.002$& $1.168\pm.009$& -&\\
   & BPL+BB&774.84/172(4.501) & $2.036\pm.003$& $3.468\pm.092$& $2.887\pm.026$&$2.058\pm.014$ &\\
   0414190801  & PL &5730.43/175(32.74) &  $1.804\pm.001$&-&-&-&\\
    & LP &558.33/174(3.21) &  $1.884\pm.001$&$-0.239\pm.003$&-& - &\\
    &  LP+BB &270.71/172(1.57) &$1.925\pm.002$ & $-0.261\pm.005$& -&$1.011\pm.038$ & 90.841\\
  & BPL &420.81/173(2.43) & $2.000\pm.003$& $1.665\pm.002$& $1.178\pm.012$& - &\\
   & BPL+BB&616.21/171(3.61) & $1.997\pm.003$& $3.001\pm.073$& $2.926\pm.033$&$2.123\pm.018$ &\\
      0414191001 & PL &3769.76/172(21.92) &  $1.803\pm.001$&-&-&-&\\
    & LP &469.77/171(2.75) &  $1.890\pm.002$&$-0.272\pm.004$&-& -&\\
    &  LP+BB &243.89/169(1.44) &$1.942\pm.004$ & $-0.300\pm.008$& - &$1.004\pm.042$&77.797\\
    & BPL &280.37/170(1.65) & $2.032\pm.005$& $1.641\pm.003$& $1.156\pm.015$& -&\\
   & BPL+BB&434.90/168(2.59) & $2.037\pm.005$ & $3.338\pm.122$& $2.727\pm.036$&$1.975\pm.020$ &\\
   0414191101  &PL&3052.38/173(17.64) &  $1.804\pm.001$&-&-&-&\\
    &  LP &301.54/172(1.75) &  $1.870\pm.001$&$-0.217\pm.003$& -&-&\\
    &  LP+BB &261.92/170(1.54) &$1.890\pm.003$ & $-0.239\pm.007$&-&$0.798\pm.066$& 12.782\\
   & BPL & 365.18/171(2.14) & $1.963\pm.004$& $1.673\pm.003$& $1.220\pm.018$& -&\\
   & BPL+BB&361.48/169(2.14) & $1.970\pm.004$& $2.835\pm.082$& $2.701\pm.038$&$2.178\pm.027$& \\
   0414191201 &PL  &7317.82/175(41.82) &  $1.662\pm.001$&-&-& -&\\
    &  LP  &514.81/174(2.96) &  $1.756\pm.001$&$-0.260\pm.003$-& -&-&\\
   & LP+BB  &299.29/172(1.74) &$1.794\pm.003$ & $-0.300\pm    .005$&-&$0.885\pm.033$& 61.568\\
   & BPL &491.43 /173(2.84 & $1.881\pm.003$& $1.517\pm.002$& $1.187\pm.011$& -&\\
   & BPL+BB& 1090.28/171(6.38) & $1.839\pm.002$& $9.080\pm1.00$& $23.291\pm1.00$&$2.111\pm.021$ &\\
   0414191301 &PL &2664.68/174(15.31) &  $1.677\pm.001$&-&-&-&\\
    & LP &352.84/173(2.04) &  $1.754\pm.002$&$-0.212\pm.004$&-&-&\\
   &  LP+BB  &295.87/171(1.73) &$1.779\pm.004$ & $-0.226\pm.008$&-&$1.051\pm.090$& 16.367\\
  & BPL &385.06/172(2.24) & $1.851\pm.005$& $1.560\pm.003$& $1.200\pm.019$& -&\\
   & BPL+BB&416.67/170(2.45) & $1.572\pm.003$& $1.571\pm1.00$& $14.681\pm1.00$&$0.103\pm.001$ &\\
   0810820101 &PL &5849.73/174(33.62)  &  $1.884\pm.001$&-&-& -&\\
    & LP &262.12/173(1.51) &  $2.001\pm.002$&$-0.325\pm.004$&-& -&1\\
    &  LP+BB &221.65/171(1.29) &$2.022\pm.004$ & $-0.341\pm.007$& -&$0.918\pm.086$& 15.519\\
   & BPL &471.43/172(2.74) & $2.110\pm.004$& $1.690\pm.003$& $1.335\pm.014$& -&\\
   & BPL+BB&699.21/170(4.11) & $2.077\pm.003$& $8.089\pm1.00$& $18.160\pm1.00$&$2.083\pm.025$ &\\
   0810821501 & PL &11246.10/176(63.89) &  $1.839\pm.001$&-&-& -&\\
   & LP &808.71/175(4.62) &  $1.974\pm.001$&$-0.360\pm.003$&-& -&\\
    &  LP+BB &301.02/173(1.74) &$2.032\pm.003$ & $-0.402\pm.006$&-&$0.985\pm.027$& 145.045\\
  & BPL &705.13 /174(4.05) & $2.146\pm.004$& $1.641\pm.002$& $1.189\pm.009$& -&\\
   & BPL+BB& 1572.52/172(9.14) & $2.090\pm.003$& $7.850\pm1.00$& $12.804\pm1.00$&$1.884\pm.013$& \\
   \hline
\end{longtable*}

\newpage

The multimodal nature of the flux histograms could be due to multiple emission or activity states in the blazar. The distributions of the flux for individual observations, however, did not feature multiple peaks. Based on the values of reduced chi-squared from the fit, either of the two distributions describes the distributions alike.
The log-normal distribution of flux has been found to be present in multi-wavelength observations of PKS 2155-304 and Mrk 421 \citep{kushwaha2017gamma, bhatta2020nature}. It is possible that the complexity of the flux distribution observed in the source can be linked to a complex interplay between the disk/corona system and the jet. For example, the flux can be dominated by disk processes at times and by jet processes at other times. This could be a possible explanation for why the source does not display a well-defined shape of the flux distribution as observed in prototypical (jet-dominated) blazars like Mrk 421.

\subsubsection{Flux-rms Relation}
While a complete account of variability in blazars is still a subject of ongoing discussion, studies focused on the possible dependence of rms on the mean flux definitely provide some clues about the variable nature of X-ray emission from blazars. The flux-rms relation implies that the variability at shorter timescales is related to or coupled with that at longer timescales. This puts strong constraints on the models of flux variability of accreting compact objects. The first evidence of a linear flux-rms relation and log-normal flux distribution in a blazar was found in the RXTE- PCA data of BL Lacertae \citep{giebels2009lognormal}. 

Similar signs were found in the Kepler light curves of the BL Lac object W2R1926+42 \citep{edelson2013kepler} as well as in the Fermi-LAT light curves of multiple blazars \citep{bhatta2020nature,Bhatta2021,kushwaha2017gamma}. More recently, hints of linear flux-rms relations were reported in the optical light curves of 12 $\gamma$-ray bright blazars including 3C 273 \cite[see][]{Bhatta2021}.

For example, theories similar to the propagating fluctuation model (\cite{lyubarskii1997flicker}), in which the longer-term fluctuations from the outer disk propagate inward along with the accretion flow and modulate the shorter timescale variability generated at the smaller radii, are favoured. In the minijets-in-a-jet model, in which mini-jets are isotropically distributed in the main jet, a linear relation between the flux and its rms might arise due to the orientation of the main jet with respect to the line of sight \citep{biteau2012minijets}.
However, the direct equivalence between flux-rms relation, multiplicative nature and log-normal flux distribution as discussed by \citet{uttley2001flux} has been disputed by \citet{scargle2020studies}.

\citet{bhattacharyya2020blazar} studied the HBL Mrk 421 using X-ray observations from AstroSAT and the Sxift/XRT instruments and reported a linear rms-flux relation. However, in our analysis, we did not observe such a linear relationship. The multimodal nature of the flux distribution suggests that the total emission is contributed by flux from various AGN components and mechanisms. Moreover, blazar variability is usually assumed to be a result of multiple processes, including particle acceleration by shocks passing through the jet or magnetic reconnection \citep{marscher1985models, joshi2014seed, joshi2010time, nalewajko2015distribution}, cooling of these particles through radiation and adiabatic expansion, as well as turbulence in the magnetic field and density \citep{marscher2013turbulent}. It remains theoretically unclear if such variability should exhibit a flux-rms relation.   

\subsection{Power Spectral Density Analysis}
In order to perform PSD analysis on the X-ray observation of the blazar 3C 273, periodograms for each of the intraday timescale observations were computed using DFT method. Consequently, the periodograms were binned in log-frequency to remove scatter and then fitted with \textit{power-law} models. The results of the analysis suggest that the periodogram can be fairly approximated by a single \textit{power-law} PSD model. The distribution of the slope indexes of the \textit{power-law} 
ranges from $-2.403$ to $2.366$. The results are comparable to the ones by similar recent work by \citet{Gowtami2022}. The results indicate that the PSD slopes tend to vary over time. Such variable slopes can result from the non-stationary variability processes taking place in general astrophysical accretion systems \citep[see, e.g.,][]{Alston2019}.
To compare the multi-wavelength slope, the slope indexes for the X-ray observations are slightly steeper than the $\gamma$-ray PSD slope of 3C 273 $\sim$ 0.77 as estimated in \citet{bhatta2020nature}. On the other hand, \citet{bhattacharyya2020blazar} shows that for 3C 273, the X-ray variability
is consistent within uncertainties across the epochs.

\subsection{Spectral Modeling: Presence of disc component}
In order to investigate the spectral properties of 3C 273, we tested various phenomenological models described in Section~\ref{sec:4} for the \textit{XMM-Newton} EPIC/PN observations spanning about 20 years. The results of the spectral fitting using different models in \textit{XSPEC} are shown in Table~\ref{table3}.

A recent multi-wavelength study of this source by \citet{fernandes2020} has shown that the accretion disc component is also required to fit the optical/X-ray data. Following this, we also included the black body component (\textit{BB}) in our further fitting process. From Table~\ref{table3} it is clear that the fit improves significantly by including the disc component in the base model LP. The presence of accretion disc is usually considered in order to explain the optical/UV emission in FSRQs \citep{jolley2009,blaes2001}.
The temperature of the disc required by the model is found to vary in the range of 0.70-1.1 keV. This indicates the varying thermal properties of the matter present in the disc. The detailed discussion of the accretion disc requires the inclusion of optical/UV data, which we plan to incorporate in the subsequent paper. 

\citet{page2004xmm} also discussed the requirement of multiple black body components to explain the varying soft excess component with temperatures ranging between around 40 and 330 eV, together with a \textit{PL}. By co-adding multiple observations, the authors demonstrated the detection of weak broad Fe-emission. In a few cases of our fitting, we noticed an indication for the presence of Fe K-$\alpha$ emission line around $6.4$ keV in the residual plot.
However, the detailed analysis of individual spectral lines is beyond the scope of our paper, and we do not discuss it here. 
We checked for the possible correlation between various spectral parameters (spectral slope, temperature, curvature parameter, mean flux) derived from the fitting. However, we were not able to establish any noticeable correlations between them.

The spectral fitting reveals the requirement of a BB component for the best fit, confirming that blazar 3C 273 is one of the few sources that exhibit a strong thermal component from the accretion disk. This could be partly due to the fact that the viewing angle to the jet of 3C 273 is oriented at a relatively larger angle of $\sim 12^{o}$ \citep{2018Natur.563..657G}, compared to $\theta < 6^{o}$ in jet-dominant blazars. As a result, the total observed emission has a significant contribution from the accretion disk, which is thermal in nature. This is also evident from the lack of correlation between gamma-ray and optical emission in \cite{Bhatta2021}. It is important to point out that we observe flux variations in the source with smaller amplitude compared to other jet-dominant sources such as Mrk 421, as indicated by the smaller values of fractional variability. This is consistent with the fact that the Doppler-boosted, beamed non-thermal emission from the jet is highly variable in comparison to the thermal emission from the accretion processes. Moreover,  any definite  trend in the flux-hardness ratio relation, e. g. linear correlation between  flux and hardness ratio and hysteresis loops,  could have been diluted due to the multi-component emission. To state a few examples of flux-spectra relation in other blazars,  FSRQ 3C 279 has been reported to show  a hysteresis loop in the colour-magnitude diagram \citep{2007ApJ...670..968B}. Similarly, spectral hysteresis loops, both clockwise and anti-clockwise, have been detected in X-ray observations of Mrk 421 \citep[e.g., see][]{2017ApJ...834....2A, 2009A&A...501..879T}.
\section{Conclusions}\label{sec:6}
We have performed spectral and timing studies of the bright blazar 3C 273 using $\sim$20 years of \textit{XMM-Newton} archival data. The fractional variability and power spectral density analysis techniques were used to examine the source's X-ray variability. X-ray continuum modelling was carried out using the \textit{XSPEC} and tested for various known models. The results obtained from these general analyses conclude our work as follows:
\begin{itemize}
\item The X-ray continuum is best represented by the log-parabolic model added with the disc emission. This result corroborates with the conclusions of other papers in the published literature \citep{page2004, fernandes2020,jolley2009,blaes2001}. This suggests that the observed X-ray emission is a combination of the jet and accretion processes.
\item The flux distributions of the individual observations follow the normal and log-normal distribution for most of the cases. However, the overall flux distribution does not follow the same trend but features multiple modes. The results point out to a scenario where the emission is contributed from various components of the jet and the disk.
\item The fractional variability and variability amplitude was found to be moderately variable throughout the observation period.  This might be attributed to the dominance of the disk component, where the variability of disk dilutes that originating from the jet.

\item The hardness ratio plot did not show any signs of a hysteresis loop, and we could not establish any notable correlation between the hardness ratio and the flux state. The observed emission, most likely contributed by the complex processes in the disk/jet system, does not reveal a definite trend as we would expect if the emission was produced by a one-zone model of blazars. 
\end{itemize}
\section*{Acknowledgements}
We thank the anonymous referee for their detailed and thoughtful comments, which have helped improve the manuscript. DG acknowledges the support of the Polish National Science Centre through the grant 2020/39 / B / ST9 / 01398. RP acknowledges that this work was supported by the Research Centre for Theoretical Physics and Astrophysics, Institute of Physics, Silesian University in Opava 
and the GA{\v C}R \mbox{23-07043S} project. ND acknowledges support from Tribhuvan University through grant HERP DLI-7B. TPA acknowledges the partial support of the Polish National Science Center's
grant No. 2021/41/B/ST9/04110.

\bibliography{sample631}{}

\begin{thebibliography}{}
\expandafter\ifx\csname natexlab\endcsname\relax\def\natexlab#1{#1}\fi
\providecommand{\url}[1]{\href{#1}{#1}}
\providecommand{\dodoi}[1]{doi:~\href{http://doi.org/#1}{\nolinkurl{#1}}}
\providecommand{\doeprint}[1]{\href{http://ascl.net/#1}{\nolinkurl{http://ascl.net/#1}}}
\providecommand{\doarXiv}[1]{\href{https://arxiv.org/abs/#1}{\nolinkurl{https://arxiv.org/abs/#1}}}

\bibitem[{{Abdo} {et~al.}(2010){Abdo}, {Ackermann}, {Ajello}, {Allafort},
  {Antolini}, {Atwood}, {Axelsson}, {Baldini}, {Ballet}, {Barbiellini},
  {Bastieri}, {Baughman}, {Bechtol}, {Bellazzini}, {Belli}, {Berenji},
  {Bisello}, {Blandford}, {Bloom}, {Bonamente}, {Bonnell}, {Borgland},
  {Bouvier}, {Bregeon}, {Brez}, {Brigida}, {Bruel}, {Burnett}, {Busetto},
  {Buson}, {Caliandro}, {Cameron}, {Campana}, {Canadas}, {Caraveo}, {Carrigan},
  {Casandjian}, {Cavazzuti}, {Ceccanti}, {Cecchi}, {{\c{C}}elik}, {Charles},
  {Chekhtman}, {Cheung}, {Chiang}, {Cillis}, {Ciprini}, {Claus},
  {Cohen-Tanugi}, {Conrad}, {Corbet}, {Davis}, {DeKlotz}, {den Hartog},
  {Dermer}, {de Angelis}, {de Luca}, {de Palma}, {Digel}, {Dormody}, {Silva},
  {Drell}, {Dubois}, {Dumora}, {Fabiani}, {Farnier}, {Favuzzi}, {Fegan},
  {Ferrara}, {Focke}, {Fortin}, {Frailis}, {Fukazawa}, {Funk}, {Fusco},
  {Gargano}, {Gasparrini}, {Gehrels}, {Germani}, {Giavitto}, {Giebels},
  {Giglietto}, {Giommi}, {Giordano}, {Giroletti}, {Glanzman}, {Godfrey},
  {Grenier}, {Grondin}, {Grove}, {Guillemot}, {Guiriec}, {Gustafsson},
  {Hadasch}, {Hanabata}, {Harding}, {Hayashida}, {Hays}, {Healey}, {Hill},
  {Horan}, {Hughes}, {Iafrate}, {J{\'o}hannesson}, {Johnson}, {Johnson},
  {Johnson}, {Johnson}, {Kamae}, {Katagiri}, {Kataoka}, {Kawai}, {Kerr},
  {Kn{\"o}dlseder}, {Kocevski}, {Kuss}, {Lande}, {Landriu}, {Latronico}, {Lee},
  {Lemoine-Goumard}, {Lionetto}, {Llena Garde}, {Longo}, {Loparco}, {Lott},
  {Lovellette}, {Lubrano}, {Madejski}, {Makeev}, {Marangelli}, {Marelli},
  {Massaro}, {Mazziotta}, {McConville}, {McEnery}, {Michelson}, {Minuti},
  {Mitthumsiri}, {Mizuno}, {Moiseev}, {Mongelli}, {Monte}, {Monzani},
  {Moretti}, {Morselli}, {Moskalenko}, {Murgia}, {Nakajima}, {Nakamori},
  {Naumann-Godo}, {Nolan}, {Norris}, {Nuss}, {Ohno}, {Ohsugi}, {Omodei},
  {Orlando}, {Ormes}, {Ozaki}, {Paccagnella}, {Paneque}, {Panetta}, {Parent},
  {Pelassa}, {Pepe}, {Pesce-Rollins}, {Pinchera}, {Piron}, {Porter}, {Poupard},
  {Rain{\`o}}, {Rando}, {Ray}, {Razzano}, {Razzaque}, {Rea}, {Reimer},
  {Reimer}, {Reposeur}, {Ripken}, {Ritz}, {Rochester}, {Rodriguez}, {Romani},
  {Roth}, {Sadrozinski}, {Salvetti}, {Sanchez}, {Sander}, {Saz Parkinson},
  {Scargle}, {Schalk}, {Scolieri}, {Sgr{\`o}}, {Shaw}, {Siskind}, {Smith},
  {Smith}, {Spandre}, {Spinelli}, {Starck}, {Stephens}, {Striani}, {Strickman},
  {Strong}, {Suson}, {Tajima}, {Takahashi}, {Takahashi}, {Tanaka}, {Thayer},
  {Thayer}, {Thompson}, {Tibaldo}, {Tibolla}, {Tinebra}, {Torres}, {Tosti},
  {Tramacere}, {Uchiyama}, {Usher}, {Van Etten}, {Vasileiou}, {Vilchez},
  {Vitale}, {Waite}, {Wallace}, {Wang}, {Watters}, {Winer}, {Wood}, {Yang},
  {Ylinen}, {Ziegler}, \& {Fermi LAT Collaboration}}]{Abdo2010}
{Abdo}, A.~A., {Ackermann}, M., {Ajello}, M., {et~al.} 2010, \apjs, 188, 405,
  \dodoi{10.1088/0067-0049/188/2/405}

\bibitem[{{Abeysekara} {et~al.}(2017){Abeysekara}, {Archambault}, {Archer},
  {Benbow}, {Bird}, {Buchovecky}, {Buckley}, {Bugaev}, {Cardenzana}, {Cerruti},
  {Chen}, {Ciupik}, {Connolly}, {Cui}, {Eisch}, {Falcone}, {Feng}, {Finley},
  {Fleischhack}, {Flinders}, {Fortson}, {Furniss}, {Griffin},
  {H{\r{a}}kansson}, {Hanna}, {Hervet}, {Holder}, {Humensky}, {H{\"u}tten},
  {Kaaret}, {Kar}, {Kertzman}, {Kieda}, {Krause}, {Kumar}, {Lang}, {Maier},
  {McArthur}, {McCann}, {Meagher}, {Moriarty}, {Mukherjee}, {Nieto}, {O'Brien},
  {Ong}, {Otte}, {Park}, {Pelassa}, {Pohl}, {Popkow}, {Pueschel}, {Ragan},
  {Reynolds}, {Richards}, {Roache}, {Sadeh}, {Santander}, {Sembroski},
  {Shahinyan}, {Staszak}, {Telezhinsky}, {Tucci}, {Tyler}, {Wakely},
  {Weinstein}, {Wilhelm}, {Williams}, {VERITAS Collaboration}, {Ahnen},
  {Ansoldi}, {Antonelli}, {Antoranz}, {Arcaro}, {Babic}, {Banerjee}, {Bangale},
  {Barres de Almeida}, {Barrio}, {Becerra Gonz{\'a}lez}, {Bednarek},
  {Bernardini}, {Berti}, {Biasuzzi}, {Biland}, {Blanch}, {Bonnefoy}, {Bonnoli},
  {Borracci}, {Bretz}, {Carosi}, {Carosi}, {Chatterjee}, {Colin}, {Colombo},
  {Contreras}, {Cortina}, {Covino}, {Cumani}, {Da Vela}, {Dazzi}, {De Angelis},
  {De Lotto}, {de O{\~n}a Wilhelmi}, {Di Pierro}, {Doert}, {Dom{\'\i}nguez},
  {Dominis Prester}, {Dorner}, {Doro}, {Einecke}, {Eisenacher Glawion},
  {Elsaesser}, {Engelkemeier}, {Fallah Ramazani}, {Fern{\'a}ndez-Barral},
  {Fidalgo}, {Fonseca}, {Font}, {Fruck}, {Galindo}, {Garc{\'\i}a L{\'o}pez},
  {Garczarczyk}, {Gaug}, {Giammaria}, {Godinovi{\'c}}, {Gora}, {Guberman},
  {Hadasch}, {Hahn}, {Hassan}, {Hayashida}, {Herrera}, {Hose}, {Hrupec},
  {Hughes}, {Idec}, {Kodani}, {Konno}, {Kubo}, {Kushida}, {Lelas}, {Lindfors},
  {Lombardi}, {Longo}, {L{\'o}pez}, {L{\'o}pez-Coto}, {Majumdar}, {Makariev},
  {Mallot}, {Maneva}, {Manganaro}, {Mannheim}, {Maraschi}, {Marcote},
  {Mariotti}, {Mart{\'\i}nez}, {Mazin}, {Menzel}, {Mirzoyan}, {Moralejo},
  {Moretti}, {Nakajima}, {Neustroev}, {Niedzwiecki}, {Nievas Rosillo},
  {Nilsson}, {Nishijima}, {Noda}, {Nogu{\'e}s}, {N{\"o}the}, {Paiano},
  {Palacio}, {Palatiello}, {Paneque}, {Paoletti}, {Paredes}, {Paredes-Fortuny},
  {Pedaletti}, {Peresano}, {Perri}, {Persic}, {Poutanen}, {Prada Moroni},
  {Prandini}, {Puljak}, {Garcia}, {Reichardt}, {Rhode}, {Rib{\'o}}, {Rico},
  {Saito}, {Satalecka}, {Schroeder}, {Schweizer}, {Shore}, {Sillanp{\"a}{\"a}},
  {Sitarek}, {Snidaric}, {Sobczynska}, {Stamerra}, {Strzys}, {Suri{\'c}},
  {Takalo}, {Tavecchio}, {Temnikov}, {Terzi{\'c}}, {Tescaro}, {Teshima},
  {Torres}, {Torres-Alb{\`a}}, {Toyama}, {Treves}, {Vanzo}, {Vazquez Acosta},
  {Vovk}, {Ward}, {Will}, {Wu}, {Zanin}, {MAGIC Collaboration}, {Hovatta}, {de
  la Calle Perez}, {Smith}, {Racero}, \& {Balokovi{\'c}}}]{2017ApJ...834....2A}
{Abeysekara}, A.~U., {Archambault}, S., {Archer}, A., {et~al.} 2017, \apj, 834,
  2, \dodoi{10.3847/1538-4357/834/1/2}

\bibitem[{Aharonian(2000)}]{aharonian2000tev}
Aharonian, F. 2000, New Astronomy, 5, 377

\bibitem[{{Alston} {et~al.}(2019){Alston}, {Fabian}, {Buisson}, {Kara},
  {Parker}, {Lohfink}, {Uttley}, {Wilkins}, {Pinto}, {De Marco}, {Cackett},
  {Middleton}, {Walton}, {Reynolds}, {Jiang}, {Gallo}, {Zogbhi}, {Miniutti},
  {Dovciak}, \& {Young}}]{Alston2019}
{Alston}, W.~N., {Fabian}, A.~C., {Buisson}, D.~J.~K., {et~al.} 2019, \mnras,
  482, 2088, \dodoi{10.1093/mnras/sty2527}

\bibitem[{{Arnaud}(1996)}]{xspec1996}
{Arnaud}, K.~A. 1996, in Astronomical Society of the Pacific Conference Series,
  Vol. 101, Astronomical Data Analysis Software and Systems V, ed. G.~H.
  {Jacoby} \& J.~{Barnes}, 17

\bibitem[{{Bhatta}(2019)}]{Bhatta2019}
{Bhatta}, G. 2019, \mnras, 487, 3990, \dodoi{10.1093/mnras/stz1482}

\bibitem[{{Bhatta}(2021)}]{Bhatta2021}
---. 2021, \apj, 923, 7, \dodoi{10.3847/1538-4357/ac2819}

\bibitem[{Bhatta \& Dhital(2020)}]{bhatta2020nature}
Bhatta, G., \& Dhital, N. 2020, The Astrophysical Journal, 891, 120

\bibitem[{Bhatta \& Webb(2018)}]{bhatta2018microvariability}
Bhatta, G., \& Webb, J.~R. 2018, Galaxies, 6, 2

\bibitem[{{Bhatta} {et~al.}(2015){Bhatta}, {Goyal}, {Ostrowski}, {Stawarz},
  {Akitaya}, {Arkharov}, {Bachev}, {Ben{\'\i}tez}, {Borman}, {Carosati},
  {Cason}, {Damljanovic}, {Dhalla}, {Frasca}, {Hu}, {Itoh}, {Jorstad},
  {Jableka}, {Kawabata}, {Klimanov}, {Kurtanidze}, {Larionov}, {Laurence},
  {Leto}, {Markowitz}, {Marscher}, {Moody}, {Moritani}, {Ohlert}, {Di Paola},
  {Raiteri}, {Rizzi}, {Sadun}, {Sasada}, {Sergeev}, {Strigachev}, {Takaki},
  {Troitsky}, {Ui}, {Villata}, {Vince}, {Webb}, {Yoshida}, {Zola}, \&
  {Hiriart}}]{Bhatta2015}
{Bhatta}, G., {Goyal}, A., {Ostrowski}, M., {et~al.} 2015, \apjl, 809, L27,
  \dodoi{10.1088/2041-8205/809/2/L27}

\bibitem[{{Bhatta} {et~al.}(2016){Bhatta}, {Stawarz}, {Ostrowski}, {Markowitz},
  {Akitaya}, {Arkharov}, {Bachev}, {Ben{\'\i}tez}, {Borman}, {Carosati},
  {Cason}, {Chanishvili}, {Damljanovic}, {Dhalla}, {Frasca}, {Hiriart}, {Hu},
  {Itoh}, {Jableka}, {Jorstad}, {Jovanovic}, {Kawabata}, {Klimanov},
  {Kurtanidze}, {Larionov}, {Laurence}, {Leto}, {Marscher}, {Moody},
  {Moritani}, {Ohlert}, {Di Paola}, {Raiteri}, {Rizzi}, {Sadun}, {Sasada},
  {Sergeev}, {Strigachev}, {Takaki}, {Troitsky}, {Ui}, {Villata}, {Vince},
  {Webb}, {Yoshida}, \& {Zola}}]{Bhatta2016}
{Bhatta}, G., {Stawarz}, {\L}., {Ostrowski}, M., {et~al.} 2016, \apj, 831, 92,
  \dodoi{10.3847/0004-637X/831/1/92}

\bibitem[{Bhattacharyya {et~al.}(2020)Bhattacharyya, Ghosh, Chatterjee, \&
  Das}]{bhattacharyya2020blazar}
Bhattacharyya, S., Ghosh, R., Chatterjee, R., \& Das, N. 2020, The
  Astrophysical Journal, 897, 25

\bibitem[{Biteau \& Giebels(2012)}]{biteau2012minijets}
Biteau, J., \& Giebels, B. 2012, Astronomy \& Astrophysics, 548, A123

\bibitem[{{Blaes} {et~al.}(2001){Blaes}, {Hubeny}, {Agol}, \&
  {Krolik}}]{blaes2001}
{Blaes}, O., {Hubeny}, I., {Agol}, E., \& {Krolik}, J.~H. 2001, \apj, 563, 560,
  \dodoi{10.1086/324045}

\bibitem[{{B{\"o}ttcher} {et~al.}(2007){B{\"o}ttcher}, {Basu}, {Joshi},
  {Villata}, {Arai}, {Aryan}, {Asfandiyarov}, {Bach}, {Bachev}, {Berduygin},
  {Blaek}, {Buemi}, {Castro-Tirado}, {de Ugarte Postigo}, {Frasca}, {Fuhrmann},
  {Hagen-Thorn}, {Henson}, {Hovatta}, {Hudec}, {Ibrahimov}, {Ishii},
  {Ivanidze}, {Jel{\'\i}nek}, {Kamada}, {Kapanadze}, {Katsuura}, {Kotaka},
  {Kovalev}, {Kovalev}, {Kub{\'a}nek}, {Kurosaki}, {Kurtanidze},
  {L{\"a}hteenm{\"a}ki}, {Lanteri}, {Larionov}, {Larionova}, {Lee}, {Leto},
  {Lindfors}, {Marilli}, {Marshall}, {Miller}, {Mingaliev}, {Mirabal},
  {Mizoguchi}, {Nakamura}, {Nieppola}, {Nikolashvili}, {Nilsson}, {Nishiyama},
  {Ohlert}, {Osterman}, {Pak}, {Pasanen}, {Peters}, {Pursimo}, {Raiteri},
  {Robertson}, {Robertson}, {Ryle}, {Sadakane}, {Sadun}, {Sigua}, {Sohn},
  {Strigachev}, {Sumitomo}, {Takalo}, {Tamesue}, {Tanaka}, {Thorstensen},
  {Tosti}, {Trigilio}, {Umana}, {Vennes}, {Vitek}, {Volvach}, {Webb},
  {Yamanaka}, \& {Yim}}]{2007ApJ...670..968B}
{B{\"o}ttcher}, M., {Basu}, S., {Joshi}, M., {et~al.} 2007, \apj, 670, 968,
  \dodoi{10.1086/522583}

\bibitem[{Cerruti {et~al.}(2015)Cerruti, Zech, Boisson, \&
  Inoue}]{cerruti2015hadronic}
Cerruti, M., Zech, A., Boisson, C., \& Inoue, S. 2015, Monthly Notices of the
  Royal Astronomical Society, 448, 910

\bibitem[{Courvoisier(1998)}]{courvoisier1998bright}
Courvoisier, T. J.-L. 1998, The Astronomy and Astrophysics Review, 9, 1

\bibitem[{Courvoisier {et~al.}(1987)Courvoisier, Turner, Robson, Gear,
  Staubert, Blecha, Bouchet, Falomo, Valtonen, \&
  Terasranta}]{courvoisier1987radio}
Courvoisier, T.-L., Turner, M., Robson, E., {et~al.} 1987, Astronomy and
  Astrophysics, 176, 197

\bibitem[{Courvoisier {et~al.}(2003)Courvoisier, Beckmann, Bourban, Chenevez,
  Chernyakova, Deluit, Favre, Grindlay, Lund, O'Brien,
  {et~al.}}]{courvoisier2003simultaneous}
Courvoisier, T.-L., Beckmann, V., Bourban, G., {et~al.} 2003, arXiv preprint
  astro-ph/0308212

\bibitem[{Dermer \& Schlickeiser(1993)}]{dermer1993model}
Dermer, C.~D., \& Schlickeiser, R. 1993, The Astrophysical Journal, 416, 458

\bibitem[{Dickey \& Lockman(1990)}]{dickey1990hi}
Dickey, J.~M., \& Lockman, F.~J. 1990, Annual review of astronomy and
  astrophysics, 28, 215

\bibitem[{Edelson {et~al.}(2013)Edelson, Mushotzky, Vaughan, Scargle, Gandhi,
  Malkan, \& Baumgartner}]{edelson2013kepler}
Edelson, R., Mushotzky, R., Vaughan, S., {et~al.} 2013, The Astrophysical
  Journal, 766, 16

\bibitem[{Edge {et~al.}(1959)Edge, Shakeshaft, McAdam, Baldwin, \&
  Archer}]{edge1959survey}
Edge, D., Shakeshaft, J., McAdam, W., Baldwin, J., \& Archer, S. 1959, Memoirs
  of the Royal Astronomical Society, 68, 37

\bibitem[{Falomo {et~al.}(2014)Falomo, Pian, \& Treves}]{falomo2014optical}
Falomo, R., Pian, E., \& Treves, A. 2014, The Astronomy and Astrophysics
  Review, 22, 73

\bibitem[{{Fan} {et~al.}(2016){Fan}, {Yang}, {Liu}, {Luo}, {Lin}, {Yuan},
  {Xiao}, {Zhou}, {Hua}, \& {Pei}}]{Fan2016}
{Fan}, J.~H., {Yang}, J.~H., {Liu}, Y., {et~al.} 2016, \apjs, 226, 20,
  \dodoi{10.3847/0067-0049/226/2/20}

\bibitem[{Fernandes {et~al.}(2020)Fernandes, Patiño-Álvarez, Chavushyan,
  Schlegel, \& Valdés}]{fernandes2020}
Fernandes, S., Patiño-Álvarez, V.~M., Chavushyan, V., Schlegel, E.~M., \&
  Valdés, J.~R. 2020, Monthly Notices of the Royal Astronomical Society, 497,
  2066, \dodoi{10.1093/mnras/staa2013}

\bibitem[{Foschini {et~al.}(2006)Foschini, Ghisellini, Raiteri, Tavecchio,
  Villata, Maraschi, Pian, Tagliaferri, Di~Cocco, \&
  Malaguti}]{foschini2006xmm}
Foschini, L., Ghisellini, G., Raiteri, C., {et~al.} 2006, Astronomy \&
  Astrophysics, 453, 829

\bibitem[{{Giannios} {et~al.}(2009){Giannios}, {Uzdensky}, \&
  {Begelman}}]{2009MNRAS.395L..29G}
{Giannios}, D., {Uzdensky}, D.~A., \& {Begelman}, M.~C. 2009, \mnras, 395, L29,
  \dodoi{10.1111/j.1745-3933.2009.00635.x}

\bibitem[{Giebels \& Degrange(2009)}]{giebels2009lognormal}
Giebels, B., \& Degrange, B. 2009, Astronomy \& Astrophysics, 503, 797

\bibitem[{{Gowtami} {et~al.}(2022){Gowtami}, {Gaur}, {Gupta}, {Wiita}, {Liao},
  \& {Ward}}]{Gowtami2022}
{Gowtami}, G.~S.~P., {Gaur}, H., {Gupta}, A.~C., {et~al.} 2022, \mnras,
  \dodoi{10.1093/mnras/stac286}

\bibitem[{{Gravity Collaboration} {et~al.}(2018){Gravity Collaboration},
  {Sturm}, {Dexter}, {Pfuhl}, {Stock}, {Davies}, {Lutz}, {Cl{\'e}net},
  {Eckart}, {Eisenhauer}, {Genzel}, {Gratadour}, {H{\"o}nig}, {Kishimoto},
  {Lacour}, {Millour}, {Netzer}, {Perrin}, {Peterson}, {Petrucci}, {Rouan},
  {Waisberg}, {Woillez}, {Amorim}, {Brandner}, {F{\"o}rster Schreiber},
  {Garcia}, {Gillessen}, {Ott}, {Paumard}, {Perraut}, {Scheithauer},
  {Straubmeier}, {Tacconi}, \& {Widmann}}]{2018Natur.563..657G}
{Gravity Collaboration}, {Sturm}, E., {Dexter}, J., {et~al.} 2018, \nat, 563,
  657, \dodoi{10.1038/s41586-018-0731-9}

\bibitem[{Haardt {et~al.}(2008)Haardt, Fossati, Grandi, Celotti, Pian,
  Ghisellini, Malizia, Maraschi, Paciesas, Raiteri,
  {et~al.}}]{haardt2008hidden}
Haardt, F., Fossati, G., Grandi, P., {et~al.} 2008, ASTROPHYSICS

\bibitem[{{Jolley} {et~al.}(2009){Jolley}, {Kuncic}, {Bicknell}, \&
  {Wagner}}]{jolley2009}
{Jolley}, E.~J.~D., {Kuncic}, Z., {Bicknell}, G.~V., \& {Wagner}, S. 2009,
  \mnras, 400, 1521, \dodoi{10.1111/j.1365-2966.2009.15554.x}

\bibitem[{Jorstad {et~al.}(2012)Jorstad, Marscher, Smith, Larionov, Agudo,
  G{\'o}mez, Casadio, Molina, \& Gurwell}]{jorstad2012parsec}
Jorstad, S., Marscher, A., Smith, P., {et~al.} 2012, in International Journal
  of Modern Physics: Conference Series, Vol.~8, World Scientific, 356--359

\bibitem[{Jorstad {et~al.}(2001)Jorstad, Marscher, Mattox, Wehrle, Bloom, \&
  Yurchenko}]{jorstad2001multiepoch}
Jorstad, S.~G., Marscher, A.~P., Mattox, J.~R., {et~al.} 2001, The
  Astrophysical Journal Supplement Series, 134, 181

\bibitem[{Joshi \& Boettcher(2010)}]{joshi2010time}
Joshi, M., \& Boettcher, M. 2010, The Astrophysical Journal, 727, 21

\bibitem[{Joshi {et~al.}(2014)Joshi, Marscher, \& Boettcher}]{joshi2014seed}
Joshi, M., Marscher, A.~P., \& Boettcher, M. 2014, The Astrophysical Journal,
  785, 132

\bibitem[{Kalita {et~al.}(2015)Kalita, Gupta, Wiita, Bhagwan, \&
  Duorah}]{kalita2015multiband}
Kalita, N., Gupta, A.~C., Wiita, P.~J., Bhagwan, J., \& Duorah, K. 2015,
  Monthly Notices of the Royal Astronomical Society, 451, 1356

\bibitem[{Kalita {et~al.}(2017)Kalita, Gupta, Wiita, Dewangan, \&
  Duorah}]{kalita2017origin}
Kalita, N., Gupta, A.~C., Wiita, P.~J., Dewangan, G.~C., \& Duorah, K. 2017,
  Monthly Notices of the Royal Astronomical Society, 469, 3824

\bibitem[{{Kirk} {et~al.}(1998){Kirk}, {Rieger}, \&
  {Mastichiadis}}]{1998A&A...333..452K}
{Kirk}, J.~G., {Rieger}, F.~M., \& {Mastichiadis}, A. 1998, \aap, 333, 452,
  \dodoi{10.48550/arXiv.astro-ph/9801265}

\bibitem[{Kushwaha {et~al.}(2017)Kushwaha, Sinha, Misra, Singh, \&
  Dal~Pino}]{kushwaha2017gamma}
Kushwaha, P., Sinha, A., Misra, R., Singh, K., \& Dal~Pino, E. d.~G. 2017, The
  Astrophysical Journal, 849, 138

\bibitem[{Leach {et~al.}(1995)Leach, McHardy, \&
  Papadakis}]{leach1995constraining}
Leach, C., McHardy, I., \& Papadakis, I. 1995, Monthly Notices of the Royal
  Astronomical Society, 272, 221

\bibitem[{Lyubarskii(1997)}]{lyubarskii1997flicker}
Lyubarskii, Y.~E. 1997, Monthly Notices of the Royal Astronomical Society, 292,
  679

\bibitem[{Maraschi {et~al.}(1992)Maraschi, Ghisellini, \&
  Celotti}]{maraschi1992jet}
Maraschi, L., Ghisellini, G., \& Celotti, A. 1992, The Astrophysical Journal,
  397, L5

\bibitem[{Marscher(2008)}]{marscher2008core}
Marscher, A.~P. 2008, in Extragalactic Jets: Theory and Observation from Radio
  to Gamma Ray, Vol. 386, 437

\bibitem[{Marscher(2013)}]{marscher2013turbulent}
Marscher, A.~P. 2013, The Astrophysical Journal, 780, 87

\bibitem[{Marscher \& Gear(1985)}]{marscher1985models}
Marscher, A.~P., \& Gear, W.~K. 1985, The Astrophysical Journal, 298, 114

\bibitem[{{Massaro} {et~al.}(2004){Massaro}, {Perri}, {Giommi}, \&
  {Nesci}}]{2004A&A...413..489M}
{Massaro}, E., {Perri}, M., {Giommi}, P., \& {Nesci}, R. 2004, \aap, 413, 489,
  \dodoi{10.1051/0004-6361:20031558}

\bibitem[{{Massaro} {et~al.}(2006){Massaro}, {Tramacere}, {Perri}, {Giommi}, \&
  {Tosti}}]{2006A&A...448..861M}
{Massaro}, E., {Tramacere}, A., {Perri}, M., {Giommi}, P., \& {Tosti}, G. 2006,
  \aap, 448, 861, \dodoi{10.1051/0004-6361:20053644}

\bibitem[{McHardy {et~al.}(2007)McHardy, Lawson, Newsam, Marscher, Sokolov,
  Urry, \& Wehrle}]{mchardy2007simultaneous}
McHardy, I., Lawson, A., Newsam, A., {et~al.} 2007, Monthly Notices of the
  Royal Astronomical Society, 375, 1521

\bibitem[{{Mohorian} {et~al.}(2022){Mohorian}, {Bhatta}, {Adhikari}, {Dhital},
  {P{\'a}nis}, {Dinesh}, {Chaudhary}, {Bachchan}, \&
  {Stuchl{\'\i}k}}]{mohorian2022}
{Mohorian}, M., {Bhatta}, G., {Adhikari}, T.~P., {et~al.} 2022, \mnras, 510,
  5280, \dodoi{10.1093/mnras/stab3738}

\bibitem[{M{\"u}cke \& Protheroe(2001)}]{mucke2001proton}
M{\"u}cke, A., \& Protheroe, R. 2001, Astroparticle Physics, 15, 121

\bibitem[{M{\"u}cke {et~al.}(2003)M{\"u}cke, Protheroe, Engel, Rachen, \&
  Stanev}]{mucke2003bl}
M{\"u}cke, A., Protheroe, R., Engel, R., Rachen, J., \& Stanev, T. 2003,
  Astroparticle Physics, 18, 593

\bibitem[{Nalewajko {et~al.}(2015)Nalewajko, Uzdensky, Cerutti, Werner, \&
  Begelman}]{nalewajko2015distribution}
Nalewajko, K., Uzdensky, D.~A., Cerutti, B., Werner, G.~R., \& Begelman, M.~C.
  2015, The Astrophysical Journal, 815, 101

\bibitem[{Page {et~al.}(2004)Page, Turner, Done, O'Brien, Reeves, Sembay, \&
  Stuhlinger}]{page2004xmm}
Page, K.~L., Turner, M.~J., Done, C., {et~al.} 2004, Monthly Notices of the
  Royal Astronomical Society, 349, 57

\bibitem[{{Page} {et~al.}(2004){Page}, {Turner}, {Done}, {O'Brien}, {Reeves},
  {Sembay}, \& {Stuhlinger}}]{page2004}
{Page}, K.~L., {Turner}, M.~J.~L., {Done}, C., {et~al.} 2004, \mnras, 349, 57,
  \dodoi{10.1111/j.1365-2966.2004.07499.x}

\bibitem[{Paltani {et~al.}(1998)Paltani, Courvoisier, \&
  Walter}]{paltani1998blue}
Paltani, S., Courvoisier, T., \& Walter, R. 1998, arXiv preprint
  astro-ph/9809113

\bibitem[{{Pavana Gowtami} {et~al.}(2022){Pavana Gowtami}, {Gaur}, {Gupta},
  {Wiita}, {Liao}, \& {Ward}}]{pavana2022}
{Pavana Gowtami}, G.~S., {Gaur}, H., {Gupta}, A.~C., {et~al.} 2022, \mnras,
  511, 3101, \dodoi{10.1093/mnras/stac286}

\bibitem[{Pietrini \& Torricelli-Ciamponi(2008)}]{pietrini2008possible}
Pietrini, P., \& Torricelli-Ciamponi, G. 2008, Astronomy \& Astrophysics, 479,
  365

\bibitem[{{Press}(1978)}]{1978ComAp...7..103P}
{Press}, W.~H. 1978, Comments on Astrophysics, 7, 103

\bibitem[{Sambruna {et~al.}(2001)Sambruna, Urry, Tavecchio, Maraschi, Scarpa,
  Chartas, \& Muxlow}]{sambruna2001chandra}
Sambruna, R.~M., Urry, C.~M., Tavecchio, F., {et~al.} 2001, The Astrophysical
  Journal Letters, 549, L161

\bibitem[{Scargle(2020)}]{scargle2020studies}
Scargle, J.~D. 2020, The Astrophysical Journal, 895, 90

\bibitem[{Schleicher {et~al.}(2019)Schleicher, Arbet-Engels, Baack, Balbo,
  Biland, Blank, Bretz, Bruegge, Bulinski, Buss,
  {et~al.}}]{schleicher2019fractional}
Schleicher, B., Arbet-Engels, A., Baack, D., {et~al.} 2019, Galaxies, 7, 62

\bibitem[{Schmidt(1963)}]{schmidt19633c}
Schmidt, M. 1963, Nature, 197, 1040

\bibitem[{Sikora {et~al.}(1994)Sikora, Begelman, \&
  Rees}]{sikora1994comptonization}
Sikora, M., Begelman, M.~C., \& Rees, M.~J. 1994, The Astrophysical Journal,
  421, 153

\bibitem[{Sillanpaa {et~al.}(1988)Sillanpaa, Haarala, Valtonen, Sundelius, \&
  Byrd}]{sillanpaa1988oj}
Sillanpaa, A., Haarala, S., Valtonen, M., Sundelius, B., \& Byrd, G. 1988, The
  Astrophysical Journal, 325, 628

\bibitem[{Soldi {et~al.}(2008)Soldi, T{\"u}rler, Paltani, Aller, Aller, Burki,
  Chernyakova, L{\"a}hteenm{\"a}ki, McHardy, Robson,
  {et~al.}}]{soldi2008multiwavelength}
Soldi, S., T{\"u}rler, M., Paltani, S., {et~al.} 2008, Astronomy \&
  Astrophysics, 486, 411

\bibitem[{Str{\"u}der {et~al.}(2001)Str{\"u}der, Briel, Dennerl, Hartmann,
  Kendziorra, Meidinger, Pfeffermann, Reppin, Aschenbach, Bornemann,
  {et~al.}}]{struder2001european}
Str{\"u}der, L., Briel, U., Dennerl, K., {et~al.} 2001, Astronomy \&
  Astrophysics, 365, L18

\bibitem[{{Tramacere} {et~al.}(2009){Tramacere}, {Giommi}, {Perri},
  {Verrecchia}, \& {Tosti}}]{2009A&A...501..879T}
{Tramacere}, A., {Giommi}, P., {Perri}, M., {Verrecchia}, F., \& {Tosti}, G.
  2009, \aap, 501, 879, \dodoi{10.1051/0004-6361/200810865}

\bibitem[{Turler {et~al.}(2000)Turler, Courvoisier, \&
  Paltani}]{turler2000modelling}
Turler, M., Courvoisier, T.-L., \& Paltani, S. 2000, arXiv preprint
  astro-ph/0008480

\bibitem[{T{\"u}rler {et~al.}(2006)T{\"u}rler, Chernyakova, Courvoisier,
  Foellmi, Aller, Aller, Kraus, Krichbaum, L{\"a}hteenm{\"a}ki, Marscher,
  {et~al.}}]{turler2006historic}
T{\"u}rler, M., Chernyakova, M., Courvoisier, T.-L., {et~al.} 2006, Astronomy
  \& Astrophysics, 451, L1

\bibitem[{Turner {et~al.}(1990)Turner, Williams, Courvoisier, Stewart, Nandra,
  Pounds, Ohashi, Makishima, Inoue, Kii, {et~al.}}]{turner1990x}
Turner, M., Williams, O., Courvoisier, T., {et~al.} 1990, Monthly Notices of
  the Royal Astronomical Society, 244, 310

\bibitem[{Urry \& Padovani(1995)}]{urry1995unified}
Urry, C.~M., \& Padovani, P. 1995, Publications of the Astronomical Society of
  the Pacific, 107, 803

\bibitem[{Uttley {et~al.}(2005)Uttley, McHardy, \& Vaughan}]{uttley2005non}
Uttley, P., McHardy, I., \& Vaughan, S. 2005, Monthly Notices of the Royal
  Astronomical Society, 359, 345

\bibitem[{Uttley \& McHardy(2001)}]{uttley2001flux}
Uttley, P., \& McHardy, I.~M. 2001, Monthly Notices of the Royal Astronomical
  Society, 323, L26

\bibitem[{Vaughan {et~al.}(2003)Vaughan, Edelson, Warwick, \&
  Uttley}]{vaughan2003characterizing}
Vaughan, S., Edelson, R., Warwick, R., \& Uttley, P. 2003, Monthly Notices of
  the Royal Astronomical Society, 345, 1271

\bibitem[{Villforth {et~al.}(2010)Villforth, Nilsson, Heidt, Takalo, Pursimo,
  Berdyugin, Lindfors, Pasanen, Winiarski, Drozdz,
  {et~al.}}]{villforth2010variability}
Villforth, C., Nilsson, K., Heidt, J., {et~al.} 2010, Monthly Notices of the
  Royal Astronomical Society, 402, 2087

\bibitem[{Wagner \& Witzel(1995)}]{wagner1995intraday}
Wagner, S., \& Witzel, A. 1995, Annual Review of Astronomy and Astrophysics,
  33, 163

\bibitem[{{Webb} {et~al.}(2021){Webb}, {Arroyave}, {Laurence}, {Revesz},
  {Bhatta}, {Hollingsworth}, {Dhalla}, {Howard}, \& {Cioffi}}]{Webb2021}
{Webb}, J.~R., {Arroyave}, V., {Laurence}, D., {et~al.} 2021, Galaxies, 9, 114,
  \dodoi{10.3390/galaxies9040114}

\end{thebibliography}
\bibliographystyle{aasjournal}

\end{document}